CENTRE DE PHYSIQUE THEORIQUE
CNRS - Luminy, Case 907
13288 Marseille Cedex 9

# ON THE UNIVERSAL CHAMSEDDINE-CONNES ACTION
# I. DETAILS OF THE ACTION COMPUTATION

**Bruno IOCHUM [1], Daniel KASTLER [2] and Thomas SCHÜCKER [1]**

**Abstract**. We give the details of the computation of the Chamseddine-Connes action by combination of a Lichnérowicz formula with the heat kernel expansion.

[1] and Université de Provence iochum@cpt.univ-mrs.fr, schucker@cpt.univ-mrs.fr
[2] and Université de la Méditerranée

**[1] *Introduction.*** Minkowskian geometry has two consequences in physics. First, starting from Coulomb's law for a static, purely electric field it generates the magnetic field together with its coupling constant $\mu_0 = 1/(\varepsilon_0\, c^2)$. Second, it changes our understanding of spacetime, no more absolute time and length. Likewise, Riemannian geometry has two consequences. First, starting from Newton's law for a static gravitational field it generates general relativity. Second, no more universal time, no more length at all. Finally, noncommutative geometry has two consequences. First, starting from a parity breaking Yang-Mills connection it generates the Higgs scalar together with its mass and spontaneous symmetry breaking [1]. Second, spacetime becomes fuzzy below a certain scale $1/\Lambda$. Within noncommutative geometry, the action of all four interactions and all 'elementary' particles consisted - so far - of three different pieces, the Dirac action for spin 1/2 fermions, the Yang-Mills-Higgs action for spin 0 and 1 bosons and the Einstein-Hilbert action for the spin 2 graviton [3]. The first two pieces are characterized by a gauge invariance, the third piece is characterized by diffeomorphism invariance. The fuzziness of spacetime originated from the first two pieces and the scale $\Lambda$ was the Higgs mass. In march this year, Alain Connes [4, 5] completed the axioms of noncommutative geometry in the sense that now there is a one-to-one correspondence between *commutative spectral triples* and *Riemannian geometries with spin*. As a by product, he suggests a natural unification of the three action pieces.

A spectral triplet consists of an associative algebra $\boldsymbol{A}$, a representation on a Hilbert space $\mathbb{H}$ classifying the fermions and a Dirac operator $\mathbb{D}$. The invariance group is simply the automorphism group of $\boldsymbol{A}$. The later is chosen to be a tensor product of the infinite dimensional, commutative algebra of differentiable functions on spacetime **M** by a matrix algebra **A** describing an internal space, $\boldsymbol{A} = C^\infty(\mathbf{M}) \otimes \mathbf{A}$. Then the automorphism group is the semi direct product of diffeomorphisms and gauge transformations. The latter are inner automorphisms. In the commutative case, $\mathbf{A} = \mathbb{C}$, there are only diffeomorphisms and the Dirac operator simply encodes the metric. If **A** is noncommutative, for instance $\mathbf{A} = \mathbb{C} \oplus \mathbb{H} \oplus M_3(\mathbb{C})$ for the standard model, then the metric 'fluctuates', that is, it picks up additional degrees of freedom from the internal space, the Yang-Mills connection and the Higgs scalar. In physicist's language, the spectral triplet is the Dirac action of a multiplet of dynamical fermions in a background field. This background field is a fluctuating metric, consisting of so far adynamical bosons of spin 0, 1 and 2. The remaining two action pieces are obtained *exclusively* from the spectrum of the covariant Dirac operator $\mathbb{D}_A$ indexed by the quantum one-form A. These two pieces together are simply the number of eigenvalues of $|\mathbb{D}_A|$ that are smaller than $\Lambda$, i.e. tr $F(|\mathbb{D}_A|/\Lambda)$ with F the characteristic function of the unit interval. This function of $\Lambda$ can be calculated conveniently from the heat kernel expansion [6] and if F was the logarithm then we would have an old physical interpretation of this action formula, the dynamics of the bosons would be *induced* from one-loop quantum corrections with fermions circulating in the loop. With the characteristic function instead, this action is essentially Klein-Gordon together with spontaneous symmetry breaking for spin 0, Yang-Mills for spin 1 and Einstein-Hilbert for spin 2. In this approach all coupling constants are fixed leading to the well known numerical problems, the Planck mass sets the scale. The hope is that these evaporate once the fuzziness of spacetime is properly taken into account. This hope is encouraged from history. Let us recall that Maxwell's relation $c^2 = (\varepsilon_0 \mu_0)^{-1}$ relates a velocity to static coupling constants. At his time the speed of light was believed to be frame dependent. It was only by accepting the revolution of Minkowskian geometry on spacetime that this problem evaporated.

The aim of this paper is to provide a detailed description of the Chamseddine-Connes bosonic action at the tree level of the standard model. The essential ingredients of this computation are Lichnérowicz' formula to get the square of the Dirac operator and the heat kernel expansion. The reader will find an index of notations at the end of the paper.

**[2]** ***Notation and reminder***. **M** is in what follows a 4-dimensional smooth compact oriented spin manifold without boundary for which we use the following notation: $\mathbb{A} = C^\infty(\mathbf{M})$, the volume element of **M** is dv, $\mathbb{S}_\mathbf{M}$ is the spin-bundle of **M** with Clifford-module of smooth sections $(\mathbb{S}(\mathbf{M}), \gamma)$, $\nabla^\mathbf{M}$ is the Levi-Civita connexion of **M** with spin connexion $\tilde{\nabla}$ with curvature $\tilde{R}$. The Clifford module of **M** is denoted by $\mathbb{C}l(\mathbf{M})$.

Our frame is that of the real spectral triple $(\boldsymbol{A}, (\mathbb{H}, \chi, \mathbb{D}), \mathbb{J})$ of the standard model (cf. [7]), tensor product of the classical real spectral triple by the inner space real spectral triple $(\mathbf{A}, (H, \chi, D), J)$ which we recall: the algebra is the tensor product $\boldsymbol{A} = C^\infty(\mathbf{M}) \otimes \mathbf{A}$ of the space-time algebra $\mathbb{A} = C^\infty(\mathbf{M})$ by the inner-space algebra $\mathbf{A} = \mathbb{C} \oplus \mathbb{H} \oplus M_3(\mathbb{C})$ where $\mathbb{H}$ denotes here the quaternion algebra. The $\mathbb{Z}/2$-graded Hilbert space $\mathbb{H}$ is the tensor product $\mathbb{H} = L^2(\mathbb{S}_\mathbf{M}) \otimes H$, the inner space K-cycle $(H, \chi, D)$ specified by $H = \underline{H} \oplus \overline{\underline{H}}$ with:

$$
(1) \qquad \begin{cases} \chi(\xi, \bar{\eta}) = (\underline{\chi}\xi, \overline{\underline{\chi}\eta}), \\ (p, q, m)(\xi, \bar{\eta}) = ((p, q)\xi, \overline{(p, m)\eta}), \\ D(\xi, \bar{\eta}) = (\underline{D}\xi, \overline{\underline{D}\eta}), \\ J(\xi, \bar{\eta}) = (\eta, \bar{\xi}), \end{cases} \qquad \begin{cases} (p, q, m) \in \mathbf{A}, \\ (\xi, \bar{\eta}) \in H, \end{cases} ,
$$

in terms of the K-cycle $(\underline{H}, \underline{D}, \underline{\chi}) = (\underline{H}_q, \underline{D}_q, \underline{\chi}_q) \oplus (\underline{H}_l, \underline{D}_l, \underline{\chi}_l)$ (the subscripts q, resp. l stand for quark, resp. lepton), with the quark and lepton parts:

(2) $$\underline{H}_q = (\underset{u_R\ d_R}{\mathbb{C}^2_R} \oplus \underset{u_L\ d_L}{\mathbb{C}^2_L}) \otimes \underset{\text{Generations}}{\mathbb{C}^N} \otimes \underset{\text{Color}}{\mathbb{C}^3} \qquad (\dim \underline{H}_q = 12N = 36),$$

(3) $$\underline{H}_l = (\underset{e_R}{\mathbb{C}^1_R} \oplus \underset{\nu_L\ e_L}{\mathbb{C}^2_L}) \otimes \underset{\text{Generations}}{\mathbb{C}^N} \otimes \mathbb{C}^1 \qquad (\dim \underline{H}_l = 3N = 9),$$

$\mathbb{Z}/2$-graded by the (right-left) parity $\chi$, and acted upon as follows by algebraic elements (p, q, m) $\in$ **A**: the quaternion q acts in the obvious way on $\mathbb{C}^2_L$, the complex scalar p acts as the diagonal quaternion $\begin{pmatrix} \bar{p} & 0 \\ 0 & p \end{pmatrix}$ on $\mathbb{C}^2_R$, and multiplies $\mathbb{C}^1_R$ by $\bar{p}$, the 3×3 matrix m acts on $\mathbb{C}^3$ (and trivially on $\mathbb{C}^1$). Finally the generalized quark, resp. lepton Dirac operators are:

(4) $$\underline{D}_q = \begin{pmatrix} 0 & 0 & M_u^* & 0 \\ 0 & 0 & 0 & M_d^* \\ M_u & 0 & 0 & 0 \\ 0 & M_d & 0 & 0 \end{pmatrix} \begin{matrix} u_R \\ d_R \\ u_L \\ d_L \end{matrix} = \begin{pmatrix} 0 & \mathbb{M}^* \\ \mathbb{M} & 0 \end{pmatrix} \begin{matrix} R \\ L \end{matrix},$$

($u_R$, $d_R$, $u_L$, $d_L$; R, L)

(5) $$\underline{D}_l = \begin{pmatrix} 0 & 0 & M_e \\ 0 & 0 & 0 \\ M_e^* & 0 & 0 \end{pmatrix} \begin{matrix} e_R \\ \nu_L \\ e_L \end{matrix},$$

($e_R$, $\nu_L$, $e_L$)

with $M_u = M_u^*$, $M_d$, $M_e = M_e^* \in \mathrm{End}\,\mathbb{C}^N$ the fermion mass-matrices.

The tensor-product full spectal triple ($\boldsymbol{A}$, ($\mathbb{H}$, $\chi$, $\mathbb{D}$), $\mathbb{J}$) is then as follows: $\mathbb{H} = L^2(\mathbb{S}_{\mathbf{M}}) \otimes H$ is acted upon by :

- the grading involution $\chi = \gamma^5 \otimes \chi$,
- the conjugation $\mathbb{J} = C \otimes J$, tensor product of the usual charge-conjugation C of euclidean electrodynamics by the conjugation J of the spectral inner space,
- the algebra $\boldsymbol{A}$ via the representation $\pi_{\mathbb{D}} = \pi_D \otimes \pi_{D}$, tensor product of the representation $\pi_D$ of $\mathbb{A}$ on $L^2(\mathbb{S}_{\mathbf{M}})$ by the representation of **A** on *H*,
- the generalized Dirac operator $\mathbb{D} = D \otimes id_H + \gamma^5 \otimes D$, $D = D^* = i\,\gamma^\mu \tilde{\nabla}_\mu$, the Dirac operator.

The spectal triple ($\boldsymbol{A}$, ($\mathbb{H}$, $\chi$, $\mathbb{D}$), $\mathbb{J}$) is an example of a "spectral geometry" in the sense of [4] - last methamorphosis of Alain Connes' noncommutative geometry. We have correspondingly, with respect to the objects considered in the previous Yang-Mills formalism, the following shift of emphasis:

- we pass on the one hand from the (generalized) Dirac operator $\mathbb{D}$ to the covariant Dirac

operator $\mathbb{D}_A = \mathbb{D} + A + \mathbb{J}A\mathbb{J}$ of the spectral formalism, indexed by the quantum one-form A.[1] - on the other hand we use instead of the Hilbert space $\mathbb{H}$ its smooth dense subspace $\mathbb{E} = \mathbb{S}(\mathbf{M}) \otimes H$.

We now give the basic:

**[3]** ***Definition***. The **universal bosonic action of M** is defined as

(6) $$\mathrm{Tr}\, F(\frac{1}{\Lambda^2} \mathbb{D}_A{}^2),$$

where $\mathbb{D}_A = \mathbb{D} + A + \mathbb{J}A\mathbb{J}$ is the "covariant Dirac operator" of the standard model in the spectral formalism. F is a function: $\mathbb{R}_+^* \to \mathbb{R}$ such that $F(\frac{1}{\Lambda^2} \mathbb{D}_A{}^2)$ is trace-class, where $\Lambda$ is the cut-off. The universal action will be the sum of the bosonic action and the fermionic action of the form $(\psi, \mathbb{D}_A\psi)$ where $\psi$ is an element of $\mathbb{E} = \mathbb{S}(\mathbf{M}) \otimes H$.

The required trace-class property, as well as positivity of the bosonic action are achieved by choosing for F: $[0, 1] \to [0, F(0)]$ a positive smooth function decreasing from a positive value F(0) to the value 0 at 1, moreover constant in a small interval $[0, \varepsilon]$, so that all its derivatives vanish at 0 (and similarly for $F_q$ and $F_l$). In fact the most sensible choice for F is the characteristic function of the interval [0, 1].

Note that the splitting $H = \underline{H} \oplus \overline{\underline{H}}$ of the inner space induces a splitting $\mathbb{H} = \mathbb{H}_{part} \oplus \mathbb{H}_{antipart}$ into Hilbert subspaces respectively pertaining to particles and antiparticules. Since the restiction of $\mathbb{D}_A$ to $\mathbb{H}_{part}$ and $\mathbb{H}_{antipart}$ are charge conjugate to each other, they yield identical contributions to the bosonic action which will be computed using only $\mathbb{D}_A$ restricted to $\mathbb{H}_{part}$ (henceforth again denoted by $\mathbb{D}_A$; from now on, $\mathbb{E}$ will correspondingly denote the particule subbundle $\mathbb{S}(\mathbf{M}) \otimes \underline{H}$.

We shall compute the bosonic action by means of a technique based on the fact that $\mathbb{D}_A$ is a generalized Dirac operator of a smooth vector bundle **V** (with set of smooth sections $\mathbb{E}$), its square $\mathbb{D}_A{}^2$ thus being a generalized laplacian. **V** is the tensor product of the spin bundle by the inner bundle with typical fiber $\underline{H} = \underline{H}_q \oplus \underline{H}_l$. Thus rank(**V**) = 36+9 = 45. Using the classical heat kernel expansion [2], we express $F(\mathbb{D}_A{}^2)$ as an integral of heat kernels (Laplace transform), thus getting for F as the following exact expression:

(7) $$\mathrm{Tr}\, F(\frac{1}{\Lambda^2} \mathbb{D}_A{}^2) = \Lambda^4 f_0\, a_0(\mathbb{D}_A{}^2) + \Lambda^2 f_2\, a_2(\mathbb{D}_A{}^2) + f_4\, a_4(\mathbb{D}_A{}^2),$$

where $f_0 = \int_{\mathbb{R}} F(u)\, u\, du$, $f_2 = \int_{\mathbb{R}} F(u)\, du$, $f_4 = F(0)$; and $a_j(\mathbb{D}_A{}^2) = \int_M a_j(x, \mathbb{D}_A{}^2)\, dv$, with:

---

[1] In the quantum Yang-Mills approach $\mathbb{D}_A$ served to define the fermionic part of the action, however now it will also play the basic role for the formulation of the bosonic action.

(8) $a_0(x, \mathbb{D}_A{}^2) = (4\pi)^{-2}\, \mathrm{tr}_x(\mathbb{1})$

(9) $a_2(x, \mathbb{D}_A{}^2) = (4\pi)^{-2}\, \mathrm{tr}_x(\frac{1}{6}\mathbf{s}\,\mathbb{1} - E)$,

(10) $a_4(x, \mathbb{D}_A{}^2) = \frac{(4\pi)^{-2}}{360}\, \mathrm{tr}_x\{5\,\mathbf{s}^2\,\mathbb{1} - 2\,\mathbf{r}^2\,\mathbb{1} + 2\,\mathbf{R}^2\,\mathbb{1} - 60\,\mathbf{s}\,E + 180\,E^2 + 30\,\mathbb{R}_{\mu\nu}\,\mathbb{R}^{\mu\nu}\}$,

where we omitted in {} the terms 12 $\mathbf{s}_{;}{}^{\alpha}{}_{\alpha}\mathbb{1}$, - 60 $E_{;}{}^{\alpha}{}_{\alpha}$ (surface-terms by Green's theorem). Here **R**, **r**, resp. **s** are the respective Levi-Civita Riemann tensor, Ricci tensor, and scalar curvature of **M**, with $\mathbf{r}^2 = \mathbf{r}_{\mu\nu}\,\mathbf{r}^{\mu\nu}$, $\mathbf{R}^2 = \mathbf{R}_{\mu\nu\alpha\beta}\,\mathbf{R}^{\mu\nu\alpha\beta}$. And E and $\mathbb{R}$ are obtained from the following prescription: express $\mathbb{D}_A{}^2$ canonically as the sum of a connection-laplacian $\Delta^{\nabla}$ and $E \in \mathrm{End}\,\mathbb{E}$ : $\mathbb{R}$ is then the endomorphism-valued curvature 2-tensor of the connexion $\nabla$ of $\mathbb{E}$.[2]

Our next task is thus to write $\mathbb{D}_A{}^2$ as the canonical decomposition $\Delta^{\nabla} + E$: this is achieved via the two following lemmas. Lemma [4] first expresses $\mathbb{D}_A$ as the sum $D^{\nabla} + \Phi$ of the Dirac operator of a Clifford bundle plus an endomorphism anticommuting with Clifford multiplication by one-forms. Lemma [5] then easily computes the canonical decomposition of $\mathbb{D}_A{}^2$ using the twisted-Clifford-bundle version of the Lichnérowicz formula.

**[4]** ***Lemma***. (i): *The dense subspace* $\mathbb{E} = \mathbb{S}(\mathbf{M}) \otimes \underline{H}$ *of the Hilbert space* $\mathbb{H}_{\mathrm{part}}$ *is a finite-projective* $\mathbb{A}$ *-module, expressible as the tensor product*

(11) $$\mathbb{E} = \mathbb{S}(\mathbf{M}) \otimes_{\mathbb{A}} \mathbf{E} \qquad \text{with} \qquad \mathbf{E} = \mathbb{A} \otimes_{\mathbb{A}} \underline{H},$$

*becoming a Clifford module* $(\mathbb{E}, c)$ *under the* $\mathbb{Z}/2$*-grading* $\chi$ *and the Clifford action*

(12) $$c = \gamma \otimes \mathrm{id}_{\mathbf{E}},$$

*and split in a direct sum of a quarkonic and the leptonic* $\mathbb{A}$ *-module according to the decomposition* $\mathbf{E} = \mathbf{E}_q \oplus \mathbf{E}_l$, *where* $\mathbf{E}_q = \mathbb{A} \otimes H_q$ *and* $\mathbf{E}_l = \mathbb{A} \otimes H_l$

(ii): *We have* $\mathbb{D}_A = D^{\nabla} + \Phi = i\, c^{\mu}\, \nabla_{\mu} + \Phi$, *direct sum* $(\mathbb{D}_A)_q \oplus (\mathbb{D}_A)_l$ *of the quarkonic and the leptonic parts:*

(13) $$\begin{cases}(\mathbb{D}_A)_q = D^{\nabla}{}_q + \Phi_q \\ (\mathbb{D}_A)_l = D^{\nabla}{}_l + \Phi_l\end{cases} \quad \text{where} \quad \begin{cases} D^{\nabla}{}_q = i\, c^{\mu}\, \nabla_{q\mu} \\ D^{\nabla}{}_l = i\, c^{\mu}\, \nabla_{l\mu}\end{cases},$$

*where, describing* End $\mathbf{E}_q$ *as* $4\times 4$ *matrices with entries in* $\mathbb{S}(\mathbf{M}) \otimes M(\mathbb{C}^N)$ *tensorized by* $M(\mathbb{C}^3_{\mathrm{colour}})$, *resp.* End $\mathbf{E}_l$ *as* $3\times 3$ *matrices with entries in* $\mathbb{S}(\mathbf{M}) \otimes M(\mathbb{C}^N)$*:*

---

[2] Our E and **s** are Gilkey's - E and - **s** **[2]**.

- *the endomorphisms* $\mathbb{\Phi}_q$ *and* $\mathbb{\Phi}_l$ *of* $\mathbb{E}$ *respectively act on the quark and lepton subspaces as the matrices:*

$$(14)\qquad \mathbb{\Phi}_q = \begin{array}{c} \begin{array}{cccc} u_R & d_R & u_L & d_L \end{array} \\ \begin{pmatrix} 0 & 0 & \Phi^2\gamma^5\otimes M_u{}^* & -\Phi^1\gamma^5\otimes M_u{}^* \\ 0 & 0 & \Phi_1\gamma^5\otimes M_d{}^* & \Phi_2\gamma^5\otimes M_d{}^* \\ \Phi_2\gamma^5\otimes M_u & \Phi^1\gamma^5\otimes M_d & 0 & 0 \\ -\Phi_1\gamma^5\otimes M_u & \Phi^2\gamma^5\otimes M_d & 0 & 0 \end{pmatrix} \end{array} \begin{array}{c} u_R \\ d_R \\ u_L \\ d_L \end{array} \otimes \mathbb{1}_3,$$

*and*

$$(15)\qquad \mathbb{\Phi}_l = \begin{array}{c} \begin{array}{ccc} e_R & \nu_L & e_L \end{array} \\ \begin{pmatrix} 0 & \Phi_1\gamma^5\otimes M_e{}^* & \Phi_2\gamma^5\otimes M_e{}^* \\ \Phi^1\gamma^5\otimes M_e & 0 & 0 \\ \Phi^2\gamma^5\otimes M_e & 0 & 0 \end{pmatrix} \end{array} \begin{array}{c} e_R \\ \nu_L \\ e_L \end{array} ,$$

*where* $\Phi_i = \bar{\Phi}^i \in C^\infty(\mathbf{M}, \mathbb{C})$*;* [3]

- *the connection* $\mathbb{\nabla}$ *of* $\mathbb{E}$ *is the tensor-product:*

$$(16)\qquad \mathbb{\nabla} = \tilde{\nabla} \otimes id_{\mathbf{E}} + id_{\mathbb{S}(\mathbf{M})} \otimes \nabla^{\mathbf{E}},$$

*of the spin connexion* $\tilde{\nabla}$ *of* $\mathbb{S}(\mathbf{M})$ *by the connexion* $\nabla^{\mathbf{E}}$ *of* $\mathbf{E}$ *specified as follows:* $\nabla^{\mathbf{E}}$ *is the direct sum* $\nabla^{\mathbf{E}}{}_q \oplus \nabla^{\mathbf{E}}{}_l$ *of a quark and a lepton connexion acting respectively on the quark and lepton subspaces as the sum of the exterior derivative and the matrices:*

$$(17)\quad id_{\mathbb{S}(\mathbf{M})}\otimes(\nabla^{\mathbf{E}}{}_q - \partial)_\mu = - i \begin{array}{c} \begin{array}{cccc} u_R & d_R & u_L & d_L \end{array} \\ \begin{pmatrix} -\mathbf{a}_\mu\otimes\mathbb{1}_N & 0 & 0 & 0 \\ 0 & \mathbf{a}_\mu\otimes\mathbb{1}_N & 0 & 0 \\ 0 & 0 & \mathbf{b}^1{}_{1\mu}\otimes\mathbb{1}_N & \mathbf{b}^1{}_{2\mu}\otimes\mathbb{1}_N \\ 0 & 0 & \mathbf{b}^2{}_{1\mu}\otimes\mathbb{1}_N & \mathbf{b}^2{}_{2\mu}\otimes\mathbb{1}_N \end{pmatrix} \end{array} \begin{array}{c} u_R \\ d_R \\ u_L \\ d_L \end{array} \otimes \mathbb{1}_3$$

$$- i\, \mathbf{c}^0{}_\mu\, \mathbb{1}_4 \otimes \mathbb{1}_N \otimes \mathbb{1}_3 - i\, \mathbf{c}^a{}_\mu\, \mathbb{1}_4 \otimes \mathbb{1}_N \otimes \frac{\lambda_a}{2} \;,$$

*and*

---

[3] Note that $\Phi^{\bullet}{}_{\bullet} = \begin{pmatrix} \Phi_2 & \Phi^1 \\ -\Phi_1 & \Phi^2 \end{pmatrix} \in C^\infty(\mathbf{M}, \mathbb{H})$.

$$(18)\quad \mathrm{id}_{\mathbb{S}(\mathbf{M})} \otimes (\nabla^{\mathbf{E}}{}_1 - \partial)_\mu = -i \begin{pmatrix} \mathbf{a}_\mu \otimes 1\!\!1_N & 0 & 0 \\ 0 & \mathbf{b}^1{}_{1\mu} \otimes 1\!\!1_N & \mathbf{b}^1{}_{2\mu} \otimes 1\!\!1_N \\ 0 & \mathbf{b}^2{}_{1\mu} \otimes 1\!\!1_N & \mathbf{b}^2{}_{2\mu} \otimes 1\!\!1_N \end{pmatrix} \begin{matrix} e_R \\ \nu_L \\ e_L \end{matrix} \; - i\, \mathbf{a}_\mu\, 1\!\!1_3 \otimes 1\!\!1_N,$$

($e_R$, $\nu_L$, $e_L$)

Here: [4]

- **a** and $\mathbf{c}_0$ are classical U(1)-vector-potentials: $\bar{\mathbf{a}} = \mathbf{a}$, $\mathbf{c}_0 \in \Omega^1(\mathbf{M}, \mathbb{C})$ ,
- $\mathbf{b}^\cdot{}_\cdot$ is a classical SU(2)-vector-potentials: $\mathbf{b}^\cdot{}_\cdot = \begin{pmatrix} \mathbf{b}^1{}_1 = \bar{\mathbf{b}}^1{}_1 & \mathbf{b}^1{}_2 = \bar{\mathbf{b}}^2{}_1 \\ \mathbf{b}^2{}_1 & \mathbf{b}^2{}_2 = -\mathbf{b}^1{}_1 \end{pmatrix} \in \Omega^1(\mathbf{M}, i\, \mathbb{H}_{\text{traceless}})$,
- $\mathbf{c}^\cdot = (\mathbf{c}^a)_{a = 1,\dots,8}$ is a SU(3)-vector-potentials (the $\lambda_a$ are the eight Gell-Mann matrices).

Proof: $\mathbb{E} = \mathbb{S}(\mathbf{M}) \otimes H$ is the finite-projective $\mathbb{A}$-module pull-back of the $\mathbb{C}$-module by the $\mathbb{A}$-$\mathbb{C}$-bimodule $\mathbb{S}(\mathbf{M})$, obviously expressible as the tensor product of $\mathbb{A}$-modules $\mathbb{E} = \mathbb{S}(\mathbf{M}) \otimes_{\mathbb{A}} \mathbf{E}$. The action (12) of $\mathbb{C}l(\mathbf{M})$ on $\mathbb{E}$ then makes it a Clifford module ($\mathbb{E}$, c), indeed $\mathbb{E}$ is a $\mathbb{Z}/2$-graded $\mathbb{C}l(\mathbf{M})$-module owing to the Clifford relations $c^\mu c^\nu + c^\nu c^\mu = g^{\mu\nu}$.

The remaining claims follow from the matrix form of the Dirac operator $\mathbb{D}$ and the vector-potentials in the spectral standard model, which we recall:

$$(19)\quad \begin{pmatrix} D\otimes 1\!\!1_N & 0 & \gamma^5\otimes M_u^* & 0 \\ 0 & D\otimes 1\!\!1_N & 0 & \gamma^5\otimes M_d^* \\ \gamma^5\otimes M_u & 0 & D\otimes 1\!\!1_N & 0 \\ 0 & \gamma^5\otimes M_d & 0 & D\otimes 1\!\!1_N \end{pmatrix} \begin{matrix} u_R \\ d_R \\ u_L \\ d_L \end{matrix}$$

($u_R$, $d_R$, $u_L$, $d_L$)

$$\begin{pmatrix} D\otimes 1\!\!1_N & 0 & \gamma^5\otimes M_e^* \\ 0 & D\otimes 1\!\!1_N & 0 \\ \gamma^5\otimes M_e & 0 & D\otimes 1\!\!1_N \end{pmatrix} \begin{matrix} e_R \\ \nu_L \\ e_L \end{matrix}$$

($e_R$, $\nu_L$, $e_L$)

and, using the shorthand $\mathbf{a}^\cdot{}_\cdot = \begin{pmatrix} -\mathbf{a} & 0 \\ 0 & \mathbf{a} \end{pmatrix}$:

(20) $i\, A = (\mathbf{a}, \mathbf{b}^\cdot{}_\cdot, \mathbf{H}, \mathbf{H}', \mathbf{c}_0, \mathbf{c}^\cdot) =$

[4] Following the physicists' usage we multiply by i our connexion one-forms to make them self-adjoint (vector potentials). Note that a quaternion is antihermitean iff it is traceless.

$u_R$ $d_R$ $u_L$ $d_L$ $e_R$ $\nu_L$ $e_L$ $\bar{u}_R$ $\bar{d}_R$ $\bar{u}_L$ $\bar{d}_L$ $\bar{e}_R$ $\bar{\nu}_L$ $\bar{e}_L$

$$\begin{pmatrix} \gamma(\mathbf{a}^{\cdot}.)\otimes\mathbb{1}_N & \mathbb{M}^*(\gamma^5\underline{\mathbf{H}}'\otimes\mathbb{1}_N) \\ (\gamma^5\underline{\mathbf{H}}\otimes\mathbb{1}_N)\mathbb{M} & \gamma(\mathbf{b}^{\cdot}.)\otimes\mathbb{1}_N \end{pmatrix}$$

Leptonic reduction
of latter

$$\gamma(\mathbf{c}_0)\mathbb{1}_2\otimes\mathbb{1}_N\otimes\mathbb{1}_3+\gamma(\mathbf{c}^a)\mathbb{1}_2\otimes\mathbb{1}_N\otimes\frac{\lambda_a}{2}$$

$$\gamma(\mathbf{a})\mathbb{1}_3\otimes\mathbb{1}_N$$

**[5]** ***Lemma***. *Let* **V** *be a smooth vector-bundle over* **M**, *with* $\mathbb{A}$*-module of smooth sections* **E**, *provided with a connexion* $\nabla^{\mathbf{E}}$ *and consider the twisted Clifford-bundle* $\mathbb{S}_{\mathbf{M}}\otimes\mathbf{V}$, $c = \gamma\otimes \mathrm{id}_{\mathbf{V}}$ *with Clifford-module* $(\mathbb{E}, c)$ *of smooth sections*

(21) $$\mathbb{E} = \mathbb{S}(\mathbf{M})\otimes\mathbf{E},$$

*and Clifford-connexion*

(22) $$\mathbb{\nabla} = \tilde{\nabla}\otimes \mathrm{id}_{\mathbf{E}} + \mathrm{id}_{\mathbb{S}(\mathbf{M})}\otimes\nabla^{\mathbf{E}},$$

*giving rise to the Dirac operator*

(23) $$\mathbb{D}^{\mathbb{\nabla}} = i\, c^\mu\, \mathbb{\nabla}_\mu \quad \text{where} \quad c^\mu = \gamma^\mu\otimes \mathrm{id}_{\mathbf{E}}.$$

*Then, given* $\Phi\in \mathrm{End}_{\mathbb{A}}(\mathbb{E})$ *anticommuting with all* $c^\mu$, *the generalized Dirac operator*

(24) $$\mathbb{D}_\Phi = i\, c^\mu\, \mathbb{\nabla}_\mu + \Phi,$$

*has the square*

(25) $$(\mathbb{D}_\Phi)^2 = \Delta^{\mathbb{\nabla}} + E,$$

*where* $\Delta^{\mathbb{\nabla}}$ *is the connexion-laplacian of* $\mathbb{\nabla}$*:*

(26) $$\Delta^{\mathbb{\nabla}} = -\, g^{\mu\nu}\,(\mathbb{\nabla}_\mu\,\mathbb{\nabla}_\nu - \Gamma^\alpha_{\mu\nu}\,\mathbb{\nabla}_\alpha),$$

*and* $E\in \mathrm{End}_{\mathbb{A}}(\mathbb{E})$ *is given by:*

(27) $$E = \frac{1}{4}\,\mathbf{s}\,\mathbb{1} - \frac{1}{2}\,c(R^{\mathbf{E}}) + ic^\mu\,[\mathbb{\nabla}_\mu, \Phi] + \Phi^2 \quad \text{with} \quad c(R^{\mathbf{E}}) = -\,\gamma^\mu\gamma^\nu\otimes R^{\mathbf{E}}(e_\mu, e_\nu),$$

*where* **s** *is the scalar curvature of* **M** *and* $R^{\mathbf{E}}$ *is the curvature of* $\nabla^{\mathbf{E}}$.

(Note that (25) is the canonical form of the generalized laplacian $(\mathbb{D}_{\Phi})^2$ as the sum of a connexion-laplacian and an endomorphism and observe that $[\nabla_\mu, \Phi]$ lies in $\mathrm{End}_{\mathbb{A}}(\mathbb{E})$ as the commutator of a $\partial_\mu$-derivation and a 0-derivation).

<u>Proof</u>: We have

$$(28)\quad (\mathbb{D}_{\Phi})^2 = (i\, c^\mu \nabla_\mu + \Phi)\,(i\, c^\nu \nabla_\nu + \Phi) = -\, c^\mu \nabla_\mu c^\nu \nabla_\nu + i\, c^\mu \nabla_\mu \Phi + i\, \Phi\, c^\mu \nabla_\mu + \Phi^2$$

$$= \mathbb{D}^{\nabla 2} + i\, c^\mu [\nabla_\mu, \Phi] + \Phi^2 = \Delta^\nabla + \frac{1}{4}\mathbf{s}\, \mathbb{1} - \frac{1}{2} c(R^{\mathbf{E}}) + i\, c^\mu [\nabla_\mu, \Phi] + \Phi^2,$$

where we plugged in the Lichnérowicz formula for the square of $\mathbb{D}^\nabla$:

$$(29)\qquad \mathbb{D}^{\nabla 2} = \Delta^\nabla + \frac{1}{4}\mathbf{s}\, \mathbb{1} - \frac{1}{2} c(R^{\mathbf{E}}).$$

**[6]** ***Lemma***. (computation of $\Phi^2$ and $\Phi^4$): (i): *We have* $\Phi^2 = \Phi_q{}^2 \oplus \Phi_l{}^2$, *with*

$$(30)\qquad \Phi_q{}^2 = |\Phi|^2 \begin{array}{c} \begin{array}{cccc} u_R & d_R & u_L & d_L \end{array} \\ \begin{pmatrix} \mathbb{1}\otimes M_u M_u{}^* & 0 & 0 & 0 \\ 0 & \mathbb{1}\otimes M_d M_d{}^* & 0 & 0 \\ 0 & 0 & \mathbb{1}\otimes M_u{}^* M_u & 0 \\ 0 & 0 & 0 & \mathbb{1}\otimes M_d{}^* M_d \end{pmatrix} \end{array} \begin{array}{c} u_R \\ d_R \\ u_L \\ d_L \end{array} \otimes \mathbb{1}_3,$$

*where* $|\Phi|^2 = \Phi_1\Phi^1 + \Phi_2\Phi^2$, *and*

$$(31)\qquad \Phi_l{}^2 = \begin{array}{c} \begin{array}{ccc} e_R & \nu_L & e_L \end{array} \\ \begin{pmatrix} |\Phi|^2\mathbb{1}\otimes M_e{}^* M_e & 0 & 0 \\ 0 & \Phi_1\Phi^1\mathbb{1}\otimes M_e M_e{}^* & \Phi_2\Phi^1\mathbb{1}\otimes M_e M_e{}^* \\ 0 & \Phi_1\Phi^2\mathbb{1}\otimes M_e M_e{}^* & \Phi_2\Phi^2\mathbb{1}\otimes M_e M_e{}^* \end{pmatrix} \end{array} \begin{array}{c} e_R \\ \nu_L \\ e_L \end{array}$$

*whence, with* $\mu_u = M_u\, M_u{}^*$, $\mu_d = M_d\, M_d{}^*$, *and* $\mu_e = M_e\, M_e{}^*$:

$$(32)\qquad \begin{cases} \mathrm{tr}_x(\Phi_q{}^2) = 8\, A_q\, |\Phi|^2 & \text{with} \quad A_q = 3\, \mathrm{tr}_N\, (\mu_u + \mu_d) \\ \mathrm{tr}_x(\Phi_l{}^2) = 8\, A_l\, |\Phi|^2 & \text{with} \quad A_l = \mathrm{tr}_N\, \mu_e \\ \mathrm{tr}_x(\Phi^2) = 8\, A\, |\Phi|^2 & \text{with} \quad A = \mathrm{tr}_N\, [3(\mu_u + \mu_d) + \mu_e]. \end{cases}$$

(ii): *We have* $\Phi^4 = \Phi_q{}^4 \oplus \Phi_l{}^4$, *with*

$$(33)\quad \Phi_q{}^4 = |\Phi|^4 \begin{array}{c} \begin{array}{cccc} u_R & d_R & u_L & d_L \end{array} \\ \begin{pmatrix} \mathbb{1}\otimes(M_u M_u{}^*)^2 & 0 & 0 & 0 \\ 0 & \mathbb{1}\otimes(M_d M_d{}^*)^2 & 0 & 0 \\ 0 & 0 & \mathbb{1}\otimes(M_u{}^* M_u)^2 & 0 \\ 0 & 0 & 0 & \mathbb{1}\otimes(M_d{}^* M_d)^2 \end{pmatrix} \end{array} \begin{array}{c} u_R \\ d_R \\ u_L \\ d_L \end{array} \otimes \mathbb{1}_3,$$

*and*

$$
\text{(34)}\qquad \mathbb{\Phi}_l^4 = \begin{array}{c} \\ \\ \\ \\ \end{array}
\begin{array}{ccc} e_R & \nu_L & e_L \\ \end{array}
$$

$$
\mathbb{\Phi}_l^4 = \begin{pmatrix} |\Phi|^4\mathbb{1}\otimes(M_e{}^*M_e)^2 & 0 & 0 \\ 0 & |\Phi|^2\Phi_1\Phi^1\mathbb{1}\otimes(M_eM_e{}^*)^2 & |\Phi|^2\Phi_2\Phi^1\mathbb{1}\otimes(M_eM_e{}^*)^2 \\ 0 & |\Phi|^2\Phi_1\Phi^2\mathbb{1}\otimes(M_eM_e{}^*)^2 & |\Phi|^2\Phi_2\Phi^2\mathbb{1}\otimes(M_eM_e{}^*)^2 \end{pmatrix}\begin{matrix} e_R \\ \nu_L \\ e_L \end{matrix}\ ,
$$

*whence*

$$
\text{(35)}\qquad \begin{cases} tr_X(\mathbb{\Phi}_q{}^4) = 8\ B_q\ |\Phi|^4 & \text{with} \quad B_q = 3\ tr_N(\mu_u{}^2 + \mu_d{}^2) \\ tr_X(\mathbb{\Phi}_l{}^4) = 8\ B_l\ |\Phi|^4 & \text{with} \quad B_l = tr_N(\mu_e{}^2) \\ tr_X(\mathbb{\Phi}^4) = 8\ B\ |\Phi|^4 & \text{with} \quad B = tr_N[3(\mu_u{}^2 + \mu_d{}^2) + \mu_e]. \end{cases}
$$

<u>Proof</u>: (i): we have $\mathbb{\Phi}_q = \begin{pmatrix} 0 & \Phi^{\cdot}{}_{\cdot}{}^* \\ \Phi^{\cdot}{}_{\cdot} & 0 \end{pmatrix} \otimes \mathbb{1}_3$, with:

$$
\text{(36)}\qquad \Phi^{\cdot}{}_{\cdot}{}^* = \begin{pmatrix} \Phi^2\gamma^5\otimes M_u{}^* & -\Phi^1\gamma^5\otimes M_u{}^* \\ \Phi_1\gamma^5\otimes M_d{}^* & \Phi_2\gamma^5\otimes M_d{}^* \end{pmatrix},\ \Phi^{\cdot}{}_{\cdot} = \begin{pmatrix} \Phi_2\gamma^5\otimes M_u & \Phi^1\gamma^5\otimes M_d \\ -\Phi_1\gamma^5\otimes M_u & \Phi^2\gamma^5\otimes M_d \end{pmatrix},
$$

thus:

$$
\text{(37)}\qquad \Phi^{\cdot}{}_{\cdot}{}^*\Phi^{\cdot}{}_{\cdot} = \begin{pmatrix} (\Phi^2\Phi_2+\Phi^1\Phi_1)\otimes M_u{}^*M_u & 0 \\ 0 & (\Phi_1\Phi^1+\Phi_2\Phi^2)\otimes M_d{}^*M_d \end{pmatrix}
$$

$$
= \begin{pmatrix} |\Phi|^2\mathbb{1}\otimes M_uM_u{}^* & 0 \\ 0 & |\Phi|^2\mathbb{1}\otimes M_dM_d{}^* \end{pmatrix},
$$

and

$$
\text{(38)}\qquad \Phi^{\cdot}{}_{\cdot}\Phi^{\cdot}{}_{\cdot}{}^* = \begin{pmatrix} |\Phi|^2\mathbb{1}\otimes M_u{}^*M_u & 0 \\ 0 & |\Phi|^2\mathbb{1}\otimes M_d{}^*M_d \end{pmatrix},
$$

and thus:

$$
\text{(39)}\qquad \mathbb{\Phi}_q{}^2 = \begin{pmatrix} \Phi^{\cdot}{}_{\cdot}{}^*\Phi^{\cdot}{}_{\cdot} & 0 \\ 0 & \Phi^{\cdot}{}_{\cdot}\Phi^{\cdot}{}_{\cdot}{}^* \end{pmatrix} \otimes \mathbb{1}_3
$$

$$
= |\Phi|^2 \begin{pmatrix} \mathbb{1}\otimes M_uM_u{}^* & 0 & 0 & 0 \\ 0 & \mathbb{1}\otimes M_dM_d{}^* & 0 & 0 \\ 0 & 0 & \mathbb{1}\otimes M_u{}^*M_u & 0 \\ 0 & 0 & 0 & \mathbb{1}\otimes M_d{}^*M_d \end{pmatrix} \otimes \mathbb{1}_3,
$$

and [5]

---

[5] $tr_N$ denotes the trace in the generation-space $\mathbb{C}^N$, N = 3.

(40) $$\mathrm{tr}_X(\mathbb{\Phi}_q{}^2) = 4\cdot 3\cdot 2\ |\Phi|^2\ \mathrm{tr}_N(\mu_u + \mu_d) = 24\ |\Phi|^2\ \mathrm{tr}_N(\mu_u + \mu_d),$$

(41) $$\mathrm{tr}_X(\mathbb{\Phi}_q{}^4) = 4\cdot 3\cdot 2\ |\Phi|^4\ \mathrm{tr}_N(\mu_u{}^2 + \mu_d{}^2) = 24\ |\Phi|^4\ \mathrm{tr}_N(\mu_u{}^2 + \mu_d{}^2),$$

(ii): We have:

(42)
$$\mathbb{\Phi}_l{}^2 = \begin{pmatrix} 0 & \Phi_1\mathbb{1}\otimes M_e{}^* & \Phi_2\gamma^5\otimes M_e{}^* \\ \Phi^1\gamma^5\otimes M_e & 0 & 0 \\ \Phi^2\gamma^5\otimes M_e & 0 & 0 \end{pmatrix}.\begin{pmatrix} 0 & \Phi_1\gamma^5\otimes M_e{}^* & \Phi_2\gamma^5\otimes M_e{}^* \\ \Phi^1\gamma^5\otimes M_e & 0 & 0 \\ \Phi^2\gamma^5\otimes M_e & 0 & 0 \end{pmatrix}$$

$$= \begin{pmatrix} |\Phi|^2\mathbb{1}\otimes M_e{}^*M_e & 0 & 0 \\ 0 & \Phi_1\Phi^1\mathbb{1}\otimes M_eM_e{}^* & \Phi_2\Phi^1\mathbb{1}\otimes M_eM_e{}^* \\ 0 & \Phi_1\Phi^2\mathbb{1}\otimes M_eM_e{}^* & \Phi_2\Phi^2\mathbb{1}\otimes M_eM_e{}^* \end{pmatrix},$$

whence:

(43) $$\mathrm{tr}_X(\mathbb{\Phi}_l{}^2) = 4\cdot 2\ |\Phi|^2\ \mathrm{tr}_N\ \mu_e = 8\ |\Phi|^2\ \mathrm{tr}_N\ \mu_e,$$

and further:

(44) $$\mathbb{\Phi}_l{}^4 = \begin{pmatrix} |\Phi|^2\mathbb{1}\otimes M_e{}^*M_e & 0 & 0 \\ 0 & \Phi_1\Phi^1\mathbb{1}\otimes M_eM_e{}^* & \Phi_2\Phi^1\mathbb{1}\otimes M_eM_e{}^* \\ 0 & \Phi_1\Phi^2\mathbb{1}\otimes M_eM_e{}^* & \Phi_2\Phi^2\mathbb{1}\otimes M_eM_e{}^* \end{pmatrix}$$

$$.\begin{pmatrix} |\Phi|^2\mathbb{1}\otimes M_e{}^*M_e & 0 & 0 \\ 0 & \Phi_1\Phi^1\mathbb{1}\otimes M_eM_e{}^* & \Phi_2\Phi^1\mathbb{1}\otimes M_eM_e{}^* \\ 0 & \Phi_1\Phi^2\mathbb{1}\otimes M_eM_e{}^* & \Phi_2\Phi^2\mathbb{1}\otimes M_eM_e{}^* \end{pmatrix}$$

$$= \begin{pmatrix} |\Phi|^4\mathbb{1}\otimes(M_e{}^*M_e)^2 & 0 & 0 \\ 0 & |\Phi|^2\Phi_1\Phi^1\mathbb{1}\otimes(M_eM_e{}^*)^2 & |\Phi|^2\Phi_2\Phi^1\mathbb{1}\otimes(M_eM_e{}^*)^2 \\ 0 & |\Phi|^2\Phi_1\Phi^2\mathbb{1}\otimes(M_eM_e{}^*)^2 & |\Phi|^2\Phi_2\Phi^2\mathbb{1}\otimes(M_eM_e{}^*)^2 \end{pmatrix}$$

and thus:

(45) $$\mathrm{tr}_X(\mathbb{\Phi}_l{}^4) = 4\ |\Phi|^2\ \left(|\Phi|^2 + \Phi_1\Phi^1 + \Phi_2\Phi^2\right)\ \mathrm{tr}_N(\mu_e{}^2) = 8\ |\Phi|^4\ \mathrm{tr}_N(\mu_e{}^2).$$

**[7]** ***Lemma***. (computation of $c^\mu\ [\mathbb{\nabla}_\mu, \mathbb{\Phi}]$ and $(c^\mu\ [\mathbb{\nabla}_\mu, \mathbb{\Phi}])^2$):

(i): *With* $\mathbf{D\Phi}^{\cdot} = \{\mathbf{D\Phi}^i = (\mathbf{D\Phi})^i\}$ *and* $\mathbf{D\Phi}_{\cdot} = \{\mathbf{D\Phi}_i = (\mathbf{D\Phi})_i\}$ *the one-forms:*

$$(46)\quad \begin{cases} \mathbf{D}\Phi^j = \mathbf{d}\Phi^j + i\,(\mathbf{a}\,\Phi^j - \mathbf{b}^j{}_k\,\Phi^k) \\ \mathbf{D}\Phi_j = \mathbf{d}\Phi_j - i\,(\mathbf{a}\,\Phi_j - \mathbf{b}^k{}_j\,\Phi_k) \end{cases},\ i=1,2,\ (\text{ i.e.} \begin{cases} \mathbf{D}\Phi^{\cdot} = \mathbf{d}\Phi^{\cdot} + i\,(\mathbf{a} - \mathbf{b}^a\,\frac{\tau_a}{2})\,\Phi^{\cdot} \\ \mathbf{D}\Phi_{\cdot} = \mathbf{d}\Phi_{\cdot} - i\,\Phi_{\cdot}\,(\mathbf{a} - \mathbf{b}^a\,\frac{\tau_a}{2}) \end{cases})$$

*we have* $c^\mu\,[\mathbb{\nabla}_\mu, \mathbb{\Phi}] = c^\mu\,[\mathbb{\nabla}_{\mu q}, \mathbb{\Phi}_q] + c^\mu\,[\mathbb{\nabla}_{\mu l}, \mathbb{\Phi}_l]$, *with*

$$(7)\quad c^\mu\,[\mathbb{\nabla}_{q\mu}, \mathbb{\Phi}_q] =$$

$$= \begin{pmatrix} 0 & 0 & \gamma(\mathbf{D}\Phi^2)\gamma^5\otimes M_u{}^* & -\gamma(\mathbf{D}\Phi^1)\gamma^5\otimes M_u{}^* \\ 0 & 0 & \gamma(\mathbf{D}\Phi_1)\gamma^5\otimes M_d{}^* & \gamma(\mathbf{D}\Phi_2)\gamma^5\otimes M_d{}^* \\ \gamma(\mathbf{D}\Phi_2)\gamma^5\otimes M_u & \gamma(\mathbf{D}\Phi^1)\gamma^5\otimes M_d & 0 & 0 \\ -\gamma(\mathbf{D}\Phi_1)\gamma^5\otimes M_u & \gamma(\mathbf{D}\Phi^2)\gamma^5\otimes M_d & 0 & 0 \end{pmatrix} \otimes \mathbb{1}_3$$

*and*

$$(48)\quad c^\mu\,[\mathbb{\nabla}_{l\mu}, \mathbb{\Phi}_l] = \begin{pmatrix} 0 & \gamma(\mathbf{D}\Phi_1)\gamma^5\otimes M_e{}^* & \gamma(\mathbf{D}\Phi_2)\gamma^5\otimes M_e{}^* \\ \gamma(\mathbf{D}\Phi^1)\gamma^5\otimes M_e & 0 & 0 \\ \gamma(\mathbf{D}\Phi^2)\gamma^5\otimes M_e & 0 & 0 \end{pmatrix},$$

*hence we have*

$$(49)\qquad tr_x\{c^\mu\,[\mathbb{\nabla}_{q\mu}, \mathbb{\Phi}_q]\} = tr_x\{c^\mu\,[\mathbb{\nabla}_{l\mu}, \mathbb{\Phi}_l]\} = tr_x\{c^\mu\,[\mathbb{\nabla}_\mu, \mathbb{\Phi}]\} = 0.$$

(ii): *We have*

$$(50)\qquad \begin{cases} tr_x(i\,c^\mu\,[\mathbb{\nabla}_{q\mu}, \mathbb{\Phi}_q])^2 = 8\,A_q\,|\mathbf{D}\Phi|^2 & \text{with } A_q = 3\,tr_N(\mu_u + \mu_d) \\ tr_x(i\,c^\mu\,[\mathbb{\nabla}_{l\mu}, \mathbb{\Phi}_l])^2 = 8\,A_l\,\mathbf{D}\Phi|^2 & \text{with } A_l = tr_N\,\mu_e \\ tr_x(i\,c^\mu\,[\mathbb{\nabla}_\mu, \mathbb{\Phi}])^2 = 8\,A\,|\mathbf{D}\Phi|^2 & \text{with } A = tr_N[3(\mu_u + \mu_d) + \mu_e]. \end{cases}$$

Proof: $\mathbb{\Phi}$ commutes with the spin-connexion one-form since the latter commutes with $\gamma^5$. It also commutes with the gluon-connexion one-forms whose matrices are diagonal with entries Clifford scalars. Thus it suffices to compute $[id_{\mathbb{S}(\mathbf{M})} \otimes (\nabla'^{\mathbf{E}}{}_q - \partial)_\mu, \mathbb{\Phi}]$, with $\nabla'^{\mathbf{E}}$ obtained from $\nabla^{\mathbf{E}}$ by deleting the gluon-connexion.

Lepton direct summand: we have:

$$(51)\quad [id_{\mathbb{S}(\mathbf{M})} \otimes (\nabla'^{\mathbf{E}}{}_l - \partial)_\mu] \cdot \mathbb{\Phi}_l =$$

$$= -i \begin{pmatrix} \mathbf{a}_\mu\otimes\mathbb{1}_N & 0 & 0 \\ 0 & \mathbf{b}^1{}_{1\mu}\otimes\mathbb{1}_N & \mathbf{b}^1{}_{2\mu}\otimes\mathbb{1}_N \\ 0 & \mathbf{b}^2{}_{1\mu}\otimes\mathbb{1}_N & \mathbf{b}^2{}_{2\mu}\otimes\mathbb{1}_N \end{pmatrix} \begin{pmatrix} 0 & \Phi_1\gamma^5\otimes M_e{}^* & \Phi_2\gamma^5\otimes M_e{}^* \\ \Phi^1\gamma^5\otimes M_e & 0 & 0 \\ \Phi^2\gamma^5\otimes M_e & 0 & 0 \end{pmatrix}$$

$$= -i \begin{pmatrix} 0 & \mathbf{a}_\mu \Phi_1 \gamma^5 \otimes M_e^* & \mathbf{a}_\mu \Phi_2 \gamma^5 \otimes M_e^* \\ (\mathbf{b}^1{}_{1\mu} \Phi^1 + \mathbf{b}^1{}_{2\mu} \Phi^2)\gamma^5 \otimes M_e & 0 & 0 \\ (\mathbf{b}^2{}_{1\mu} \Phi^1 + \mathbf{b}^2{}_{2\mu} \Phi^2)\gamma^5 \otimes M_e & 0 & 0 \end{pmatrix},$$

(52) $\mathbb{\Phi}_l \cdot [id_{\mathbb{S}(\mathbf{M})} \otimes (\nabla'^{\mathbf{E}}{}_l - \partial)_\mu] =$

$$= -i \begin{pmatrix} 0 & \Phi_1 \gamma^5 \otimes M_e^* & \Phi_2 \gamma^5 \otimes M_e^* \\ \Phi^1 \gamma^5 \otimes M_e & 0 & 0 \\ \Phi^2 \gamma^5 \otimes M_e & 0 & 0 \end{pmatrix} \begin{pmatrix} \mathbf{a}_\mu \otimes \mathbb{1}_N & 0 & 0 \\ 0 & \mathbf{b}^1{}_{1\mu} \otimes \mathbb{1}_N & \mathbf{b}^1{}_{2\mu} \otimes \mathbb{1}_N \\ 0 & \mathbf{b}^2{}_{1\mu} \otimes \mathbb{1}_N & \mathbf{b}^2{}_{2\mu} \otimes \mathbb{1}_N \end{pmatrix},$$

$$= -i \begin{pmatrix} 0 & (\Phi_1 \mathbf{b}^1{}_{1\mu} + \Phi_2 \mathbf{b}^2{}_{1\mu})\gamma^5 \otimes M_e^* & (\Phi_1 \mathbf{b}^1{}_{2\mu} + \Phi_2 \mathbf{b}^2{}_{2\mu})\gamma^5 \otimes M_e^* \\ \Phi^1 \mathbf{a}_\mu \gamma^5 \otimes M_e & 0 & 0 \\ \Phi^2 \mathbf{a}_\mu \gamma^5 \otimes M_e & 0 & 0 \end{pmatrix},$$

thus we have:

(53) $-i\, [id_{\mathbb{S}(\mathbf{M})} \otimes (\nabla^{\mathbf{E}}{}_l - \partial)_\mu, \mathbb{\Phi}_l] =$

$$\begin{pmatrix} 0 & (\mathbf{a}_\mu \Phi_1 - \Phi_1 \mathbf{b}^1{}_{1\mu} - \Phi_2 \mathbf{b}^2{}_{1\mu})\gamma^5 \otimes M_e^* & (\mathbf{a}_\mu \Phi_2 - \Phi_1 \mathbf{b}^1{}_{2\mu} - \Phi_2 \mathbf{b}^2{}_{2\mu})\gamma^5 \otimes M_e^* \\ (\mathbf{b}^1{}_{1\mu} \Phi^1 + \mathbf{b}^1{}_{2\mu} \Phi^2 - \Phi^1 \mathbf{a}_\mu)\gamma^5 \otimes M_e & 0 & 0 \\ (\mathbf{b}^2{}_{1\mu} \Phi^1 + \mathbf{b}^2{}_{2\mu} \Phi^2 - \Phi^2 \mathbf{a}_\mu)\gamma^5 \otimes M_e & 0 & 0 \end{pmatrix}$$

(54) $$[\mathbb{\nabla}_{l\mu}, \mathbb{\Phi}_l] = \begin{pmatrix} 0 & (\mathbf{D}\Phi_1)_\mu \gamma^5 \otimes M_e^* & (\mathbf{D}\Phi_2)_\mu \gamma^5 \otimes M_e^* \\ \gamma^5 (\mathbf{D}\Phi^1)_\mu \otimes M_e & 0 & 0 \\ \gamma^5 (\mathbf{D}\Phi^2)_\mu \otimes M_e & 0 & 0 \end{pmatrix},$$

thus

(55) $$c^\mu\, [\mathbb{\nabla}_{l\mu}, \mathbb{\Phi}_l] = \begin{pmatrix} 0 & \gamma(\mathbf{D}\Phi_1)\gamma^5 \otimes M_e^* & \gamma(\mathbf{D}\Phi_2)\gamma^5 \otimes M_e^* \\ \gamma(\mathbf{D}\Phi^1)\gamma^5 \otimes M_e & 0 & 0 \\ \gamma(\mathbf{D}\Phi^2)\gamma^5 \otimes M_e & 0 & 0 \end{pmatrix}$$

and thus:

(56) $(c^{\mu}\,[\mathbb{\nabla}_{l\mu}, \mathbb{\Phi}_l])^2 =$

$$\begin{pmatrix} 0 & \gamma(\mathbf{D}\Phi_1)\gamma^5\otimes M_e^* & \gamma(\mathbf{D}\Phi_2)\gamma^5\otimes M_e^* \\ \gamma(\mathbf{D}\Phi^1)\gamma^5\otimes M_e & 0 & 0 \\ \gamma(\mathbf{D}\Phi^2)\gamma^5\otimes M_e & 0 & 0 \end{pmatrix}\begin{pmatrix} 0 & \gamma(\mathbf{D}\Phi_1)\gamma^5\otimes M_e^* & \gamma(\mathbf{D}\Phi_2)\gamma^5\otimes M_e^* \\ \gamma(\mathbf{D}\Phi^1)\gamma^5\otimes M_e & 0 & 0 \\ \gamma(\mathbf{D}\Phi^2)\gamma^5\otimes M_e & 0 & 0 \end{pmatrix}$$

$$= - \begin{pmatrix} \gamma(\mathbf{D}\Phi_k)\gamma(\mathbf{D}\Phi^k)\otimes M_e^* M_e & 0 & 0 \\ 0 & \gamma(\mathbf{D}\Phi_1)\gamma(\mathbf{D}\Phi^1)\otimes M_e M_e^* & 0 \\ 0 & 0 & \gamma(\mathbf{D}\Phi_2)\gamma(\mathbf{D}\Phi^2)\otimes M_e M_e^* \end{pmatrix},$$

whence:

(57) $\mathrm{Tr}_X\{(i\, c^{\mu}\,[\mathbb{\nabla}_{l\mu}, \mathbb{\Phi}_l])^2\} = 8\, |\mathbf{D}\Phi|^2\, \mathrm{tr}_N\, \mu_e = 8\, A_l\, |\mathbf{D}\Phi|^2.$

Quark direct summand: we have:

(58) $i\;[\mathrm{id}_{\mathbb{S}(\mathbf{M})} \otimes (\nabla'^{\mathbf{E}}_q - \partial)_\mu] \cdot \mathbb{\Phi}_q =$

$$\begin{pmatrix} -\mathbf{a}_\mu\otimes\mathbb{1}_N & 0 & 0 & 0 \\ 0 & \mathbf{a}_\mu\otimes\mathbb{1}_N & 0 & 0 \\ 0 & 0 & \mathbf{b}^1{}_{1\mu}\otimes\mathbb{1}_N & \mathbf{b}^1{}_{2\mu}\otimes\mathbb{1}_N \\ 0 & 0 & \mathbf{b}^2{}_{1\mu}\otimes\mathbb{1}_N & \mathbf{b}^2{}_{2\mu}\otimes\mathbb{1}_N \end{pmatrix}\begin{pmatrix} 0 & 0 & \Phi^2\gamma^5\otimes M_u^* & -\Phi^1\gamma^5\otimes M_u^* \\ 0 & 0 & \Phi_1\gamma^5\otimes M_d^* & \Phi_2\gamma^5\otimes M_d^* \\ \Phi_2\gamma^5\otimes M_u & \Phi^1\gamma^5\otimes M_d & 0 & 0 \\ -\Phi_1\gamma^5\otimes M_u & \Phi^2\gamma^5\otimes M_d & 0 & 0 \end{pmatrix}\otimes\mathbb{1}_3,$$

$$= \begin{pmatrix} 0 & 0 & -\mathbf{a}_\mu\Phi^2\gamma^5\otimes M_u^* & \mathbf{a}_\mu\Phi^1\gamma^5\otimes M_u^* \\ 0 & 0 & \mathbf{a}_\mu\Phi_1\gamma^5\otimes M_d^* & \mathbf{a}_\mu\Phi_2\gamma^5\otimes M_d^* \\ -(\mathbf{b}^k{}_{2\mu}\Phi_k)\gamma^5\otimes M_u & (\mathbf{b}^1{}_{k\mu}\Phi^k)\gamma^5\otimes M_d & 0 & 0 \\ (\mathbf{b}^k{}_{1\mu}\Phi_k)\gamma^5\otimes M_u & (\mathbf{b}^2{}_{k\mu}\Phi^k)\gamma^5\otimes M_d & 0 & 0 \end{pmatrix}\otimes\mathbb{1}_3,$$

(59) $i\;\mathbb{\Phi}_q\cdot[\mathrm{id}_{\mathbb{S}(\mathbf{M})} \otimes (\nabla^{\mathbf{E}}_q - \partial)_\mu] =$

$$\begin{pmatrix} 0 & 0 & \Phi^2\gamma^5\otimes M_u^* & -\Phi^1\gamma^5\otimes M_u^* \\ 0 & 0 & \Phi_1\gamma^5\otimes M_d^* & \Phi_2\gamma^5\otimes M_d^* \\ \Phi_2\gamma^5\otimes M_u & \Phi^1\gamma^5\otimes M_d & 0 & 0 \\ -\Phi_1\gamma^5\otimes M_u & \Phi^2\gamma^5\otimes M_d & 0 & 0 \end{pmatrix}\begin{pmatrix} -\mathbf{a}_\mu\otimes\mathbb{1}_N & 0 & 0 & 0 \\ 0 & \mathbf{a}_\mu\otimes\mathbb{1}_N & 0 & 0 \\ 0 & 0 & \mathbf{b}^1{}_{1\mu}\otimes\mathbb{1}_N & \mathbf{b}^1{}_{2\mu}\otimes\mathbb{1}_N \\ 0 & 0 & \mathbf{b}^2{}_{1\mu}\otimes\mathbb{1}_N & \mathbf{b}^2{}_{2\mu}\otimes\mathbb{1}_N \end{pmatrix}\otimes\mathbb{1}_3$$

$$
= \begin{pmatrix} 0 & 0 & -(\mathbf{b}^2{}_{2\mu}\Phi^2+\mathbf{b}^2{}_{1\mu}\Phi^1)\gamma^5\otimes M_u^* & (\mathbf{b}^1{}_{2\mu}\Phi^2+\mathbf{b}^1{}_{1\mu}\Phi_1)\gamma^5\otimes M_u^* \\ 0 & 0 & (\mathbf{b}^1{}_{1\mu}\Phi_1+\mathbf{b}^2{}_{1\mu}\Phi_2)\gamma^5\otimes M_d^* & (\mathbf{b}^1{}_{2\mu}\Phi_1+\mathbf{b}^2{}_{2\mu}\Phi_2)\gamma^5\otimes M_d^* \\ -\mathbf{a}_\mu\Phi_2\gamma^5\otimes M_u & \mathbf{a}_\mu\Phi^1\gamma^5\otimes M_d & 0 & 0 \\ \mathbf{a}_\mu\Phi_1\gamma^5\otimes M_u & \mathbf{a}_\mu\Phi^2\gamma^5\otimes M_d & 0 & 0 \end{pmatrix} \otimes \mathbb{1}_3
$$

thus:

(60) $\quad i\,[\mathrm{id}_{\mathbb{S}(\mathbf{M})} \otimes (\nabla^{\mathbf{E}}{}_q - \partial)_\mu, \mathbb{\Phi}_q\,] =$

$$
\begin{pmatrix} 0 & 0 & -(\mathbf{a}_\mu\Phi^2-\mathbf{b}^2{}_{k\mu}\Phi^k)\gamma^5\otimes M_u^* & (\mathbf{a}_\mu\Phi^1-\mathbf{b}^1{}_{k\mu}\Phi^k)\gamma^5\otimes M_u^* \\ 0 & 0 & (\mathbf{a}_\mu\Phi_1-\mathbf{b}_{1k\mu}\Phi^k)\gamma^5\otimes M_d^* & (\mathbf{a}_\mu\Phi_2-\mathbf{b}_{2k\mu}\Phi^k)\gamma^5\otimes M_d^* \\ (\mathbf{a}_\mu\Phi_2-\mathbf{b}^k{}_{2\mu}\Phi_k)\gamma^5\otimes M_u & -(\mathbf{a}_\mu\Phi^1-\mathbf{b}^1{}_{k\mu}\Phi^k)\gamma^5\otimes M_d & 0 & 0 \\ -(\mathbf{a}_\mu\Phi_1-\mathbf{b}_{1k\mu}\Phi^k)\gamma^5\otimes M_u & -(\mathbf{a}_\mu\Phi^2-\mathbf{b}^2{}_{k\mu}\Phi^k)\gamma^5\otimes M_d & 0 & \end{pmatrix}
$$

$$
\otimes\ \mathbb{1}_3
$$

whence

$$
[\mathbb{\nabla}_{q\mu}, \mathbb{\Phi}_q] = \begin{pmatrix} 0 & 0 & \gamma^5\mathbf{D}\Phi^2\otimes M_u^* & -\gamma^5\mathbf{D}\Phi^1\otimes M_u^* \\ 0 & 0 & \gamma^5\mathbf{D}\Phi_1\otimes M_d^* & \gamma^5\mathbf{D}\Phi_2\otimes M_d^* \\ \gamma^5\mathbf{D}\Phi_2\otimes M_u & \gamma^5\mathbf{D}\Phi^1\otimes M_d & 0 & 0 \\ -\gamma^5\mathbf{D}\Phi_1\otimes M_u & \gamma^5\mathbf{D}\Phi^2\otimes M_d & 0 & 0 \end{pmatrix} \otimes \mathbb{1}_3
$$

and

$$
(61)\ c^\mu\,[\mathbb{\nabla}_{q\mu}, \mathbb{\Phi}_q] = \begin{pmatrix} 0 & 0 & \gamma(\mathbf{D}\Phi^2)\gamma^5\otimes M_u^* & -\gamma(\mathbf{D}\Phi^1)\gamma^5\otimes M_u^* \\ 0 & 0 & \gamma(\mathbf{D}\Phi_1)\gamma^5\otimes M_d^* & \gamma(\mathbf{D}\Phi_2)\gamma^5\otimes M_d^* \\ \gamma(\mathbf{D}\Phi_2)\gamma^5\otimes M_u & \gamma(\mathbf{D}\Phi^1)\gamma^5\otimes M_d & 0 & 0 \\ -\gamma(\mathbf{D}\Phi_1)\gamma^5\otimes M_u & \gamma(\mathbf{D}\Phi^2)\gamma^5\otimes M_d & 0 & 0 \end{pmatrix} \otimes \mathbb{1}_3.
$$

(62) upper corner of $([\mathbb{\nabla}_{q\mu}, \mathbb{\Phi}_q])^2 =$

$$
= \begin{pmatrix} \gamma(\mathbf{D}\Phi^2)\gamma^5\otimes M_u^* & -\gamma(\mathbf{D}\Phi^1)\gamma^5\otimes M_u^* \\ \gamma(\mathbf{D}\Phi_1)\gamma^5\otimes M_d^* & \gamma(\mathbf{D}\Phi_2)\gamma^5\otimes M_d^* \end{pmatrix} \begin{pmatrix} \gamma(\mathbf{D}\Phi_2)\gamma^5\otimes M_u & \gamma(\mathbf{D}\Phi^1)\gamma^5\otimes M_d \\ -\gamma(\mathbf{D}\Phi_1)\gamma^5\otimes M_u & \gamma(\mathbf{D}\Phi^2)\gamma^5\otimes M_d \end{pmatrix}
$$

$$
= -\begin{pmatrix} [\gamma(\mathbf{D}\Phi_2)\gamma(\mathbf{D}\Phi^2)+\gamma(\mathbf{D}\Phi_1)\gamma(\mathbf{D}\Phi^1)]\otimes M_u^*M_u & [\gamma(\mathbf{D}\Phi^2)\gamma(\mathbf{D}\Phi^1)-\gamma(\mathbf{D}\Phi^1)\gamma(\mathbf{D}\Phi^2)]\gamma\otimes M_u^*M_d \\ [\gamma(\mathbf{D}\Phi_1)\gamma(\mathbf{D}\Phi_2)-\gamma(\mathbf{D}\Phi_2)\gamma(\mathbf{D}\Phi_1)]\otimes M_d^*M_u & [\gamma(\mathbf{D}\Phi_1)\gamma(\mathbf{D}\Phi^1)+\gamma(\mathbf{D}\Phi_2)\gamma(\mathbf{D}\Phi^2)]\otimes M_d^*M_d \end{pmatrix}
$$

whence

(63) $\mathrm{tr}_x\{\text{upper corner }([\mathbb{\nabla}_{q\mu}, \mathbb{\Phi}_q])^2\} = - 4\, |\mathbf{D\Phi}|^2\, \mathrm{tr}_N(\mu_u + \mu_d)$
$$= \mathrm{tr}_x\{\text{lower corner }([\mathbb{\nabla}_{q\mu}, \mathbb{\Phi}_q])^2\},$$

thus

(64) $$\mathrm{Tr}_x\{(\, i\, c^\mu\, [\mathbb{\nabla}_{q\mu}, \mathbb{\Phi}_q])^2\} = 8\, |\mathbf{D\Phi}|^2\, \mathrm{tr}_N[3(\mu_u + \mu_d)] = 8\, A_q\, |\mathbf{D\Phi}|^2.$$

It remains us to compute $c(R)^2$ and $\mathbb{R}_{\mu\nu}\, \mathbb{R}^{\mu\nu}$:

**[8]** ***Lemma.*** (computation of $c(R^{\mathbf{E}})^2$ and $\mathbb{R}_{\mu\nu}\, \mathbb{R}^{\mu\nu}$). *Let* $(e_\mu, \varepsilon^\mu)$ *be an orthonormal frame of the tangent bundle of* **M**. (i): *We have:*

(65) $$\mathrm{tr}_x[c(R^{\mathbf{E}})^2] = - 8\, \mathrm{tr}_{\mathbf{V}}(R^{\mathbf{E}}{}_{\mu\nu}\, R^{\mathbf{E}\mu\nu})$$

(ii): *We have:*

(66) $$\mathrm{tr}_x[\mathbb{R}_{\mu\nu}\, \mathbb{R}^{\mu\nu}] = - \tfrac{1}{2}(\mathrm{rank}\mathbf{V})\, \mathbf{R}^2 + 4\, \mathrm{tr}_{\mathbf{V}}(R^{\mathbf{E}}{}_{\mu\nu}\, R^{\mathbf{E}\mu\nu}).$$

<u>Proof</u>: (i): We have, since $R^{\mathbf{E}}{}_{\mu\nu} = - R^{\mathbf{E}}{}_{\nu\mu}$:

(67) $$\mathrm{tr}_x[c(R^{\mathbf{E}})^2] = \mathrm{tr}_x[((\gamma^\mu\, \gamma^\nu \otimes R^{\mathbf{E}}(e_\mu, e_\nu))^2] = \mathrm{tr}_x[(\gamma^\mu\, \gamma^\nu\, \gamma^\alpha\, \gamma^\beta \otimes R^{\mathbf{E}}(e_\mu, e_\nu)\, R^{\mathbf{E}}(e_\alpha, e_\beta)]$$

$$= 4\, (g^{\mu\nu\alpha\beta} + g^{\mu\beta\nu\alpha} - g^{\mu\alpha\nu\beta})\, \mathrm{tr}_{\mathbf{V}}(R^{\mathbf{E}}{}_{\mu\nu}\, R^{\mathbf{E}}{}_{\alpha\beta})$$

$$= 4\, (g^{\mu\beta\nu\alpha} - g^{\mu\alpha\nu\beta})\, \mathrm{tr}_{\mathbf{V}}(R^{\mathbf{E}}{}_{\mu\nu}\, R^{\mathbf{E}}{}_{\alpha\beta}) = 4\, \mathrm{tr}_{\mathbf{V}}(R^{\mathbf{E}\beta}{}_{\nu}\, R^{\mathbf{E}\nu}{}_{\beta} - R^{\mathbf{E}\alpha}{}_{\nu}\, R^{\mathbf{E}}{}_{\alpha}{}^{\nu})$$

$$= 4\, \mathrm{tr}_{\mathbf{V}}(R^{\mathbf{E}}{}_{\beta\nu}\, R^{\mathbf{E}\nu\beta} - R^{\mathbf{E}}{}_{\alpha\nu}\, R^{\mathbf{E}\alpha\nu}) = - 8\, \mathrm{tr}_{\mathbf{V}}(R^{\mathbf{E}}{}_{\mu\nu}\, R^{\mathbf{E}\mu\nu})$$

(ii): We have:

(68) $$\mathrm{tr}_x[\mathbb{R}_{\mu\nu}\, \mathbb{R}^{\mu\nu}] = \mathrm{tr}_x[(\widetilde{R}_{\mu\nu} \otimes \mathrm{id}_{\mathbb{E}} + \mathrm{id}_{\mathbb{S}_{\mathbf{M}}} \otimes R^{\mathbf{E}}{}_{\mu\nu})(\widetilde{R}^{\mu\nu} \otimes \mathrm{id}_{\mathbb{E}} + \mathrm{id}_{\mathbb{S}_{\mathbf{M}}} \otimes R^{\mathbf{E}\mu\nu}).]$$

$$= \mathrm{tr}_x(\widetilde{R}_{\mu\nu}\, \widetilde{R}^{\mu\nu} \otimes \mathrm{id}_{\mathbf{V}}) + \mathrm{tr}_x(\mathrm{id}_{\mathbb{S}_{\mathbf{M}}} \otimes R^{\mathbf{E}}{}_{\mu\nu}\, R^{\mathbf{E}\mu\nu}) + 2\, \mathrm{tr}_x(\widetilde{R}_{\mu\nu} \otimes R^{\mathbf{E}\mu\nu})$$

$$= (\mathrm{rank}\mathbf{V})\, \mathrm{tr}_{\mathbb{S}_{\mathbf{M}}}(\widetilde{R}_{\mu\nu}\, \widetilde{R}^{\mu\nu}) + 4\, \mathrm{tr}_{\mathbf{V}}(R^{\mathbf{E}}{}_{\mu\nu}\, R^{\mathbf{E}\mu\nu})$$

$$= - \tfrac{1}{2}(\mathrm{rank}\mathbf{V})\, \mathbf{R}^2 + 4\, \mathrm{tr}_{\mathbf{V}}(R^{\mathbf{E}}{}_{\mu\nu}\, R^{\mathbf{E}\mu\nu}),$$

where we plugged the value

$$\mathrm{tr}_{\mathbb{S}_{\mathbf{M}}}(\widetilde{R}_{\mu\nu}\, \widetilde{R}^{\mu\nu}) = \tfrac{1}{16}\mathbf{R}_{\mu\nu\alpha\beta}\, \mathbf{R}^{\mu\nu}{}_{\sigma\tau}\, \mathrm{tr}_x(\gamma^\alpha\, \gamma^\beta\, \gamma^\sigma\, \gamma^\tau)$$
$$= \tfrac{1}{4}\mathbf{R}_{\mu\nu\alpha\beta}\, \mathbf{R}^{\mu\nu}{}_{\sigma\tau}\, (g^{\alpha\beta}\, g^{\sigma\tau} + g^{\alpha\tau}\, g^{\beta\sigma} - g^{\alpha\sigma}\, g^{\beta\tau})$$

$$= \frac{1}{4}(\mathbf{R}_{\mu\nu}{}^{\tau\sigma}\, \mathbf{R}^{\mu\nu}{}_{\sigma\tau} - \mathbf{R}_{\mu\nu}{}^{\sigma\tau}\, \mathbf{R}^{\mu\nu}{}_{\sigma\tau})$$
$$= -\frac{1}{2}\, \mathbf{R}_{\mu\nu}{}^{\sigma\tau}\, \mathbf{R}^{\mu\nu}{}_{\sigma\tau} = -\frac{1}{2}\mathbf{R}^2.$$

and took account of the fact that $tr_{\mathbb{S}_{\mathbf{M}}}(\widetilde{R}_{\mu\nu}) = \frac{1}{4} tr_{\mathbb{S}_{\mathbf{M}}}(\mathbf{R}_{\mu\nu\alpha\beta}\, \gamma^\alpha \gamma^\beta) = \frac{1}{4} tr_{\mathbb{S}_{\mathbf{M}}}(\mathbf{R}_{\mu\nu\alpha\beta}\, g^{\alpha\beta}) = 0$.

We now have the ingredients of the computation of our heat-kernel expansion.

<u>Computation of</u> $a_0(x, \mathbb{D}_A{}^2)$: we have $(4\pi)^2\, a_0(x, \mathbb{D}_A{}^2) = tr_x(\mathbb{1}) = 4\, \text{rank}\mathbf{V}$.

<u>Computation of</u> $a_2(x, \mathbb{D}_A{}^2)$: we have to compute $(4\pi)^2\, a_2(x, \mathbb{D}_A{}^2) = tr_x(\frac{1}{6}\mathbf{s}\, \mathbb{1} - E)$, recalling that we found $E = \frac{1}{4}\mathbf{s}\, \mathbb{1} - \frac{1}{2} c(R^{\mathbf{E}}) + i\, c^\mu\, [\mathbb{V}_\mu, \mathbb{\Phi}] + \mathbb{\Phi}^2$. With $\cong$ denoting equality under $tr_x$, we have $\frac{1}{6}\mathbf{s}\, \mathbb{1} - E \cong \frac{1}{6}\mathbf{s}\, \mathbb{1} - \frac{1}{4}\mathbf{s}\, \mathbb{1} - \mathbb{\Phi}^2$, whence

$$(4\pi)^2\, a_2(x, \mathbb{D}_A{}^2) = -\frac{\text{rank}\mathbf{V}}{3}\, \mathbf{s} - 8\, A\, |\Phi|^2.$$

(for the convenience of the reader we recall that we found:

$$(32) \begin{cases} tr_x(\mathbb{\Phi}_q{}^2) = 8A_q|\Phi|^2 \\ tr_x(\mathbb{\Phi}_l{}^2) = 8A_l|\Phi|^2 \\ tr_x(\mathbb{\Phi}^2) = 8A|\Phi|^2 \end{cases} \quad (35) \begin{cases} tr_x(\mathbb{\Phi}_q{}^4) = 8B_q|\Phi|^4 \\ tr_x(\mathbb{\Phi}_l{}^4) = 8B_l|\Phi|^4 \\ tr_x(\mathbb{\Phi}^4) = 8B|\Phi|^4 \end{cases} \quad (50) \begin{cases} tr_x(ic^\mu[\mathbb{V}_{q\mu}, \mathbb{\Phi}_q])^2 = 8A_q|\mathbf{D}\Phi|^2 \\ tr_x(ic^\mu[\mathbb{V}_{l\mu}, \mathbb{\Phi}_l])^2 = 8A_l|\mathbf{D}\Phi|^2 \\ tr_x(ic^\mu[\mathbb{V}_\mu, \mathbb{\Phi}])^2 = 8A|\mathbf{D}\Phi|^2 \end{cases} ).$$

<u>Computation of</u> $a_4(x, \mathbb{D}_A{}^2)$: We have, with the shorthands $\mathbf{r}^2 = \mathbf{r}_{\mu\nu}\, \mathbf{r}^{\mu\nu}$ and $\mathbf{R}^2 = \mathbf{R}_{\mu\nu\alpha\beta}\, \mathbf{R}^{\mu\nu\alpha\beta}$ :

(69) $360\ (4\pi)^2\, a_4(x, \mathbb{D}_A{}^2) = tr_x\{5\, \mathbf{s}^2\, \mathbb{1} - 2\, \mathbf{r}^2\, \mathbb{1} + 2\, \mathbf{R}^2\, \mathbb{1} - 60\, \mathbf{s}\, E + 180\, E^2 + 30\, \mathbb{R}_{\mu\nu}\, \mathbb{R}^{\mu\nu}\}$

$$= tr_x\{(\tfrac{5}{4}\mathbf{s}^2 - 2\, \mathbf{r}^2 + 2\, \mathbf{R}^2)\mathbb{1} + 30\, \mathbf{s}\ \mathbb{\Phi}^2 + 180\, \mathbb{\Phi}^4 + 180\, (i\, c^\mu\, [\mathbb{V}_\mu, \mathbb{\Phi}])^2$$
$$+ 45\, c(R)^2 + 30\, \mathbb{R}_{\mu\nu}\, \mathbb{R}^{\mu\nu}\}$$
$$= (\text{rank}\mathbf{V})\, (5\, \mathbf{s}^2 - 8\, \mathbf{r}^2 - 7\, \mathbf{R}^2) + 240\, A\, \mathbf{s}\, \Phi|^2 + 1440\, B\, |\Phi|^4$$
$$+ 1440\, A\, |\mathbf{D}\Phi|^2 + 120\, tr_{\mathbf{V}}(R^{\mathbf{E}}{}_{\mu\nu}\, R^{\mathbf{E}\mu\nu})$$
$$= -18\, (\text{rank}\mathbf{V})\, \mathbf{C}^2 + 240\, A\, \mathbf{s}\, |\Phi|^2 + 1440\, B\, |\Phi|^4 + 1440\, A\, |\, \mathbf{D}\Phi|^2$$
$$- 240\, tr_{\mathbf{V}}(R^{\mathbf{E}}{}_{\mu\nu}\, R^{\mathbf{E}\mu\nu}).$$

We first took account of the fact that we have:

(70) $-60\, \mathbf{s}\, E + 180\, E^2 \cong -60\, [\frac{1}{4}\mathbf{s}^2\, \mathbb{1} + \mathbf{s}\mathbb{\Phi}^2] + 45\, [\frac{1}{4}\mathbf{s}^2\, \mathbb{1} + c(R^{\mathbf{E}})^2 + 4\, (i\, c^\mu\, [\mathbb{V}_\mu, \mathbb{\Phi}])^2$

$$+ 4\, \mathbb{\Phi}^4 + 2\, \mathbf{s}\mathbb{\Phi}^2]$$
$$\cong -\frac{15}{4}\, \mathbf{s}^2\, \mathbb{1} + 30\, \mathbf{s}\, \mathbb{\Phi}^2 + 180\, \mathbb{\Phi}^4 + 180\, (i\, c^\mu\, [\mathbb{V}_\mu, \mathbb{\Phi}])^2 + 45\, c(R^{\mathbf{E}})^2,$$

neglecting the cross-terms in $E^2$ involving $c(R)$ or/and $c^\mu[\mathbb{V}_\mu, \mathbb{\Phi}]$ (these vanish under $tr_x$ : owing to $tr\, \gamma^\mu = 0$ and $tr\, \gamma^\mu \gamma^\nu = tr\, \gamma^\nu \gamma^\mu$ ). We then plugged into (69) the values (65) and (66), and finally effected the replacement:

(71) $$5\,\mathbf{s}^2 - 8\,\mathbf{r}^2 - 7\,\mathbf{R}^2 = 88\,\pi^2\chi_4 - 18\,\mathbf{C}^2,$$

where $\mathbf{C}^2 = \mathbf{R}^2 - \mathbf{r}^2 + \frac{1}{3}\mathbf{s}^2$ is the square of the Weyl tensor, and we subsequently suppress $\chi_4 = 2\,(4\pi)^{-2}\,(\mathbf{R}^2 - 4\,\mathbf{r}^2 + \mathbf{s}^2)$, since $\chi_4$ dv is the Euler form, which, by the Gauss-Bonnet theorem, does not contribute to the action. We proved:

**[9] *Lemma***. *We have, for the bundle* **V** *(the module* $\mathbb{E}$*):* [6]

(72) $$(4\pi)^2\, a_0(x, \mathbb{D}_A{}^2) = 4\,\mathrm{rank}\mathbf{V},$$

(73) $$(4\pi)^2\, a_2(x, \mathbb{D}_A{}^2) = -\,\frac{\mathrm{rank}\mathbf{V}}{3}\,\mathbf{s} - 8\,A\,|\Phi|^2$$

(74) $$(4\pi)^2\, a_4(x, \mathbb{D}_A{}^2) = -\frac{1}{20}(\mathrm{rank}\mathbf{V})\,\mathbf{C}^2 + \frac{2}{3}\,A\,\mathbf{s}|\Phi|^2 + 4\,B\,|\Phi|^4 + 4\,A\,|\mathbf{D}\Phi|^2 - \frac{2}{3}\,\mathrm{tr}_{\mathbf{V}}(R^{\mathbf{E}}{}_{\mu\nu}\,R^{\mathbf{E}\mu\nu})$$

We now write the bosonic action density, as given by (7), which we rewrite for convenience: [7]

(7) $$\mathrm{Tr}\,F(\tfrac{1}{\Lambda^2}\mathbb{D}_A{}^2) = \Lambda^4 f_0\, a_0(\mathbb{D}_A{}^2) + \Lambda^2 f_2\, a_2(\mathbb{D}_A{}^2) + f_4\, a_4(\mathbb{D}_A{}^2), \quad \begin{cases} f_0 = \int F(u)\,u\,du, \\ f_2 = \int F(u)\,du, \\ f_4 = F(0). \end{cases}$$

For rank**V** = 45, we get:

**[10] *Proposition***. *The bosonic Lagrangian* $I_B$ *is given for a general function* F *as follows:*

(75) $$(4\pi)^2 I_B = 180\,\Lambda^4 f_0 - 15\,\Lambda^2 f_2\,\mathbf{s} - 8\,\Lambda^2 A\, f_2\,|\Phi|^2 - \frac{9}{4} f_4\,\mathbf{C}^2 + \frac{2}{3} f_4\,A\,\mathbf{s}\,|\Phi|^2 + 4\,B\,f_4\,|\Phi|^4 + 4\,A\,f_4\,|\mathbf{D}\Phi|^2 - \frac{2}{3} f_4\,\mathrm{tr}_{\mathbf{V}}(R^{\mathbf{E}}{}_{\mu\nu}\,R^{\mathbf{E}\mu\nu}),$$

*and for the choice* F = *characteristic function of* [0,1] *corresponding to* $f_0 = \frac{1}{2}$, $f_2 = f_4 = 1$,

(76) $$(4\pi)^2 I_B = 90\,\Lambda^4 - 15\,\Lambda^2\,\mathbf{s} - 8\,\Lambda^2 A\,|\Phi|^2 - \frac{9}{4}\mathbf{C}^2 + \frac{2}{3}\,A\,\mathbf{s}\,|\Phi|^2 + 4\,B\,|\Phi|^4 + 4\,A\,|\mathbf{D}\Phi|^2 - \frac{2}{3}\mathrm{tr}_{\mathbf{V}}(R^{\mathbf{E}}{}_{\mu\nu}\,R^{\mathbf{E}\mu\nu}),$$

Proof: We have, with possible changes as mentioned in footnote 6:

---

[6] with results for the bundles $\mathbf{V}_q$, resp. $\mathbf{V}_l$ (the section modules $\mathbb{E}_q$, resp. $\mathbb{E}_l$) obtained by the replacements rank$\mathbf{V} \to$ rank$\mathbf{V}_q$, $A \to A_q$, $B \to B_q$ and $\mathrm{tr}_{\mathbf{V}} \to \mathrm{tr}_{\mathbf{V}_q}$, resp. rank$\mathbf{V} \to$ rank$\mathbf{V}_l$, $A \to A_l$, $B \to B_l$ and $\mathrm{tr}_{\mathbf{V}} \to \mathrm{tr}_{\mathbf{V}_l}$.

[7] At this point we keep the global expression $\frac{1}{3}\,\mathrm{tr}_{\mathbf{V}}(R_{\mu\nu}\,R^{\mu\nu})$ of the gluonic part of the action, defering its detailed description to the next proposition which also discusses the modular adjustment.

(77) $(4\pi)^2 I_B = 4\,\Lambda^4 f_0\,\mathrm{rank}\mathbf{V} - \Lambda^2 f_2\,[\frac{\mathrm{rank}\mathbf{V}}{3}\mathbf{s} + 8\,A\,|\Phi|^2\,]$

$$+ f_4\,[-\frac{1}{20}(\mathrm{rank}\mathbf{V})\,\mathbf{C}^2 + \frac{2}{3}A\,\mathbf{s}\,|\Phi|^2 + 4\,B\,|\Phi|^4 + 4\,A\,|\mathbf{D}\Phi|^2 - \frac{2}{3}\mathrm{tr}_{\mathbf{V}}(R^{\mathbf{E}}{}_{\mu\nu}\,R^{\mathbf{E}\mu\nu})]$$

$$= 4\,\Lambda^4 f_0\,\mathrm{rank}\mathbf{V} - \Lambda^2\frac{\mathrm{rank}\mathbf{V}}{3} f_2\,\mathbf{s} - \frac{\mathrm{rank}\mathbf{V}}{20} f_4\,\mathbf{C}^2 - 8\,\Lambda^2 A\,f_2|\,\Phi|^2 + \frac{2}{3} f_4\,A\,\mathbf{s}\,|\Phi|^2$$
$$+ 4\,B\,f_4\,|\Phi|^4 + 4\,A\,f_4\,\,|\mathbf{D}\Phi|^2 - \frac{2}{3} f_4\,\mathrm{tr}_{\mathbf{V}}(R^{\mathbf{E}}{}_{\mu\nu}\,R^{\mathbf{E}\mu\nu}).$$

**[11]** ***Proposition***. (computation of $\mathrm{tr}_{\mathbf{V}}(R'^{\mathbf{E}}{}_{\mu\nu}\,R'^{\mathbf{E}\mu\nu})$): *The gluonic part of the bosonic Lagrangian is as follows: with* $R'^{\mathbf{E}}{}_{\mu\nu}$ *the curvature-tensor obtained from* $R^{\mathbf{E}}{}_{\mu\nu}$ *by performing the "modular adjustment" , i.e. setting* $\mathbf{g}_0 = -\frac{1}{3}\mathbf{f}$, *one has:*

(78) $$- \mathrm{tr}_{\mathbf{V}}(R'^{\mathbf{E}}{}_{q\mu\nu}\,R'^{\mathbf{E}q\mu\nu}) = \frac{40}{3}N\,\mathbf{f}_{\mu\nu}\,\mathbf{f}^{\mu\nu} + 2N\,\mathbf{h}^s{}_{\mu\nu}\,\mathbf{h}_s{}^{\mu\nu} + 2N\,\mathbf{g}^a{}_{\mu\nu}\,\mathbf{g}_a{}^{\mu\nu}$$

(one has separately:

(79) $$- \mathrm{tr}_{\mathbf{V}_q}(R'_{q\mu\nu}\,R'^{q\mu\nu}) = \frac{22}{3}N\,\mathbf{f}_{\mu\nu}\,\mathbf{f}^{\mu\nu} + \frac{3}{2}N\,\mathbf{h}^s{}_{\mu\nu}\,\mathbf{h}_s{}^{\mu\nu} + 2N\,\mathbf{g}^a{}_{\mu\nu}\,\mathbf{g}_a{}^{\mu\nu},$$

and

(80) $$- \mathrm{tr}_{\mathbf{V}_l}(R'_{l\mu\nu}\,R'^{l\mu\nu}) = \frac{18}{3}N\,\mathbf{f}_{\mu\nu}\,\mathbf{f}^{\mu\nu} + \frac{N}{2}\mathbf{h}^s{}_{\mu\nu}\,\mathbf{h}_s{}^{\mu\nu}\ ).$$

Here:
- $\mathbf{f}$ and $\mathbf{g}_0$ are classical U(1)-curvatures: $\bar{\mathbf{f}} = \mathbf{f}$, $\mathbf{g}_0 \in \Omega(\mathbf{M}, \mathbb{C})^2$ ,
- $\mathbf{h}^{\cdot}{}_{\cdot}$ is a classical SU(2)-curvatures: $\mathbf{h}^{\cdot}{}_{\cdot} = \begin{pmatrix} \mathbf{h}^1{}_1=\bar{\mathbf{h}}^1{}_1 & \mathbf{h}^1{}_2=\bar{\mathbf{h}}^2{}_1 \\ \mathbf{h}^2{}_1 & \mathbf{h}^2{}_2=-\mathbf{h}^1{}_1 \end{pmatrix} \in \Omega^2(\mathbf{M}, i\,\mathbb{H}_{\mathrm{traceless}})$,
- $\mathbf{g}^{\cdot} = (\mathbf{g}^a)\,\frac{\lambda_a}{2}$ is a SU(3)-curvatures (the $\lambda_a$, a = 1,..., 8, are the Gell-Mann matrices).

<u>Proof</u>: In accordance with (17), (18) we have:

(81) $$i\,R^{\mathbf{E}}{}_{q\mu\nu} = \begin{matrix} & u_R & d_R & u_L & d_L & \\ & \left(\begin{matrix} -\mathbf{f}_{\mu\nu}\otimes\mathbb{1}_N \\ 0 \\ 0 \\ 0 \end{matrix}\right. & \begin{matrix} 0 \\ \mathbf{f}_{\mu\nu}\otimes\mathbb{1}_N \\ 0 \\ 0 \end{matrix} & \begin{matrix} 0 \\ 0 \\ \mathbf{h}^1{}_{1\mu\nu}\otimes\mathbb{1}_N \\ \mathbf{h}^2{}_{1\mu\nu}\otimes\mathbb{1}_N \end{matrix} & \left.\begin{matrix} 0 \\ 0 \\ \mathbf{h}^1{}_{2\mu\nu}\otimes\mathbb{1}_N \\ \mathbf{h}^2{}_{2\mu\nu}\otimes\mathbb{1}_N \end{matrix}\right) & \begin{matrix} u_R \\ d_R \\ u_L \\ d_L \end{matrix} \end{matrix} \otimes \mathbb{1}_3$$
$$+ \mathbf{g}^0{}_{\mu\nu}\,\mathbb{1}_4\otimes\mathbb{1}_N\otimes\mathbb{1}_3 + \mathbf{g}^a{}_{\mu\nu}\,\mathbb{1}_4\otimes\mathbb{1}_N\otimes\frac{\lambda_a}{2}$$

transformed by modular adjustment into:

$$(81a)\qquad i\,R'^{\mathbf{E}}_{q\mu\nu} = \begin{pmatrix} -\frac{4}{3}\mathbf{f}_{\mu\nu}\otimes 1_N & 0 & 0 & 0 \\ 0 & \frac{2}{3}\mathbf{f}_{\mu\nu}\otimes 1_N & 0 & 0 \\ 0 & 0 & \underline{\mathbf{h}}^1{}_{1\mu\nu}\otimes 1_N & \underline{\mathbf{h}}^1{}_{2\mu\nu}\otimes 1_N \\ 0 & 0 & \underline{\mathbf{h}}^2{}_{1\mu\nu}\otimes 1_N & \underline{\mathbf{h}}^2{}_{2\mu\nu}\otimes 1_N \end{pmatrix} \begin{matrix} u_R \\ d_R \\ u_L \\ d_L \end{matrix} \otimes 1_3$$

($u_R$, $d_R$, $u_L$, $d_L$)

$$+ \mathbf{g}^a{}_{\mu\nu}\, 1_4 \otimes 1_N \otimes \frac{\lambda_a}{2} \;,$$

and

$$(82)\qquad i\,R^{\mathbf{E}}_{l\mu\nu} = \begin{pmatrix} 2\mathbf{f}_{\mu\nu}\otimes 1_N & 0 & 0 \\ 0 & \bar{\mathbf{h}}^1{}_{1\mu\nu}\otimes 1_N & \bar{\mathbf{h}}^1{}_{2\mu\nu}\otimes 1_N \\ 0 & \bar{\mathbf{h}}^2{}_{1\mu\nu}\otimes 1_N & \bar{\mathbf{h}}^2{}_{2\mu\nu}\otimes 1_N \end{pmatrix} \begin{matrix} e_R \\ \nu_L \\ e_L \end{matrix}$$

($e_R$, $\nu_L$, $e_L$)

where $\underline{\mathbf{h}}^i{}_k = \mathbf{h}^i{}_k - \frac{1}{3}\,\mathbf{f}\,\delta^i{}_k$, and $\bar{\mathbf{h}} = \mathbf{h} + \mathbf{f}\,\delta^i{}_k$. We thus have after some algebra

$$(83)\quad -\,R'^{\mathbf{E}}_{q\mu\nu}\, R'^{\mathbf{E}}_q{}^{\mu\nu} =$$

$$\begin{pmatrix} \frac{16}{9}\mathbf{f}_{\mu\nu}\mathbf{f}^{\mu\nu}\otimes 1_N & 0 & 0 & 0 \\ 0 & \frac{4}{9}\mathbf{f}_{\mu\nu}\mathbf{f}^{\mu\nu}\otimes 1_N & 0 & 0 \\ 0 & 0 & \underline{\mathbf{h}}^1{}_{i\mu\nu}\underline{\mathbf{h}}^i{}_1{}^{\mu\nu}\otimes 1_N & \underline{\mathbf{h}}^1{}_{i\mu\nu}\underline{\mathbf{h}}^i{}_2{}^{\mu\nu}\otimes 1_N \\ 0 & 0 & \underline{\mathbf{h}}^2{}_{i\mu\nu}\underline{\mathbf{h}}^i{}_1{}^{\mu\nu}\otimes 1_N & \underline{\mathbf{h}}^2{}_{i\mu\nu}\underline{\mathbf{h}}^i{}_2{}^{\mu\nu}\otimes 1_N \end{pmatrix} \begin{matrix} u_R \\ d_R \\ u_L \\ d_L \end{matrix} \otimes 1_3$$

($u_R$, $d_R$, $u_L$, $d_L$)

$$+ \mathbf{g}^a{}_{\mu\nu}\, \mathbf{g}_a{}^{\mu\nu}\, 1_4 \otimes 1_N \otimes 1_3 \;,$$

*and*

$$(84)\quad -\,R'^{\mathbf{E}}_{l\mu\nu}\, R'^{\mathbf{E}}_l{}^{\mu\nu} = \begin{pmatrix} 4\mathbf{f}_{\mu\nu}\mathbf{f}^{\mu\nu}\otimes 1_N & 0 & 0 \\ 0 & \bar{\mathbf{h}}^1{}_{i\mu\nu}\bar{\mathbf{h}}^i{}_1{}^{\mu\nu}\otimes 1_N & \bar{\mathbf{h}}^1{}_{i\mu\nu}\bar{\mathbf{h}}^i{}_2{}^{\mu\nu}\otimes 1_N \\ 0 & \bar{\mathbf{h}}^2{}_{i\mu\nu}\bar{\mathbf{h}}^i{}_1{}^{\mu\nu}\otimes 1_N & \bar{\mathbf{h}}^2{}_{i\mu\nu}\bar{\mathbf{h}}^i{}_2{}^{\mu\nu}\otimes 1_N \end{pmatrix} \begin{matrix} e_R \\ \nu_L \\ e_L \end{matrix} \;.$$

($e_R$, $\nu_L$, $e_L$)

Quark contribution to $tr_{\mathbf{V}}(R'^{\mathbf{E}}_{\mu\nu}\, R'^{\mathbf{E}\mu\nu})$: for the computation tensor-product traces $Tr_p \otimes Tr_3$, we shall use the following elementary fact: we have, due to the tracelessness of the $\lambda_a$ and the relation $Tr(\lambda_a\, \lambda_b) = 2\,\delta_{ab}$:

$$(85)\qquad \{Tr_p \otimes Tr_3\}[M \otimes 1_3 + N^a \otimes \frac{\lambda_a}{2}]^2 = 3\;\; Tr_p M^2 + \frac{1}{2} Tr_p(N^a\, N_a), \qquad M, N^a \in M_p(\mathbb{C}).$$

From this follows that we have no mixed electoweak-chromodynamics terms, and that the gluonic contribution is:

(86) $$2N\, \mathbf{g}^a{}_{\mu\nu}\, \mathbf{g}_a{}^{\mu\nu}.$$

Noting that we have, by the fact that the $\mathbf{h}^i{}_k$ are traceless:

(87) $$\underline{\mathbf{h}}^i{}_{k\mu\nu}\, \underline{\mathbf{h}}^k{}_i{}^{\mu\nu} = \mathbf{h}^i{}_{k\mu\nu}\, \mathbf{h}^k{}_i{}^{\mu\nu} + \frac{2}{9}\, \mathbf{f}_{\mu\nu}\, \mathbf{f}^{\mu\nu},$$

we get the electoweak contribution:

(88) $$3N\,(\frac{16}{9}\, \mathbf{f}_{\mu\nu}\, \mathbf{f}^{\mu\nu} + \frac{4}{9}\, \mathbf{f}_{\mu\nu}\, \mathbf{f}^{\mu\nu} + \mathbf{h}^i{}_{k\mu\nu}\, \mathbf{h}^k{}_i{}^{\mu\nu} + \frac{2}{9}\, \mathbf{f}_{\mu\nu}\, \mathbf{f}^{\mu\nu}) = \frac{22}{3}\, N\, \mathbf{f}_{\mu\nu}\, \mathbf{f}^{\mu\nu} + \frac{3}{2}\, N\, \mathbf{h}^s{}_{\mu\nu}\, \mathbf{h}_s{}^{\mu\nu}$$

(we used the fact that

(89) $$\mathbf{h}^s{}_{\mu\nu}\; \mathbf{h}_s{}^{\mu\nu} = 2\, \mathbf{h}^i{}_{k\mu\nu}\, \mathbf{h}^k{}_i{}^{\mu\nu}\; ).$$

Lepton contribution to $tr_{\mathbf{V}}(R'^{\mathbf{E}}{}_{\mu\nu}\, R'^{\mathbf{E}\mu\nu})$: using

(90) $$\bar{\mathbf{h}}^i{}_{k\mu\nu}\, \bar{\mathbf{h}}^k{}_i{}^{\mu\nu} = \mathbf{h}^i{}_{k\mu\nu}\, \mathbf{h}^k{}_i{}^{\mu\nu} + 2\, \mathbf{f}_{\mu\nu}\, \mathbf{f}^{\mu\nu},$$

we get,, using again (89)

(91) $$N\,(4\, \mathbf{f}_{\mu\nu}\, \mathbf{f}^{\mu\nu} + \mathbf{h}^i{}_{k\mu\nu}\, \mathbf{h}^k{}_i{}^{\mu\nu} + 2\, \mathbf{f}_{\mu\nu}\, \mathbf{f}^{\mu\nu}) = 6N\, \mathbf{f}_{\mu\nu}\, \mathbf{f}^{\mu\nu} + \frac{1}{2}\, N\, \mathbf{h}^s{}_{\mu\nu}\, \mathbf{h}_s{}^{\mu\nu}.$$

Concluding, we have:

**[12] *Theorem***. *For* F *the characteristic function of the interval* [0, 1], *the bosonic Lagrangian is given as follows:* [8]

(92) $$(4\pi)^2\, I_B = 90\, \Lambda^4 - 15\, \Lambda^2\, \mathbf{s} - 8\, \Lambda^2\, A\, |\Phi|^2 - \frac{9}{4}\, \mathbf{C}^2 + \frac{2}{3}\, A\, \mathbf{s}\, |\Phi|^2 + 4\, B\, |\Phi|^4 + 4\, A\, |\mathbf{D}\Phi|^2 + \frac{80}{9}\, N\, \mathbf{f}_{\mu\nu}\, \mathbf{f}^{\mu\nu} + \frac{4}{3}\, N\, \mathbf{h}^s{}_{\mu\nu}\, \mathbf{h}_s{}^{\mu\nu} + \frac{4}{3}\, N\, \mathbf{g}^a{}_{\mu\nu}\, \mathbf{g}_a{}^{\mu\nu},$$

*where* $\Lambda$ *is the cut-off and:*

(93) $$\begin{cases} A = tr_N[3(M_u{}^*M_u + M_d{}^*M_d) + M_e{}^*M_e], \\ B = tr_N[3(M_u{}^*M_uM_u{}^*M_u + M_d{}^*M_dM_d{}^*M_d) + M_e{}^*M_eM_e{}^*M_e]. \end{cases}$$

---

[8] after modular adjustment.

This result agrees with the Chamseddine-Connes paper [6] [9] with the following correspondence of notation:

| | [6] | the present paper |
|---|---|---|
| | H | $\mathbf{\Phi}$ |
| | $g_{01}\, B_\mu\,,\; g_{01}\, B_{\mu\nu}$ | $-2\,\mathbf{a}_\mu\,,\; -2\,\mathbf{f}_\mu$ |
| | $g_{02}\, A^{\cdot}_{\mu}\,,\; g_{02}\, F^{\cdot}_{\mu\nu}$ | $\mathbf{b}^{\cdot}_{\mu}\,,\; \mathbf{h}^{\cdot}_{\mu\nu}$ |
| (94) | $g_{03}\, G^{\cdot}_{\mu}\,,\; g_{02}\, C^{\cdot}_{\mu\nu}$ | $\mathbf{c}^{\cdot}_{\mu}\,,\; \mathbf{g}^{\cdot}_{\mu\nu}$ |
| | R | $-\mathbf{s}$ |
| | $y^2$ | $\frac{1}{3}A$ |
| | $z^2$ | $\frac{1}{3}B$ |

**[13]** ***Remark***. The modular adjustment $R^{\mathbf{E}}_{\mu\nu} \to R'^{\mathbf{E}}_{\mu\nu}$ that we have performed was before taking $tr_{\mathbf{V}}$ of the square. But in fact one gets the same result if one first compute $tr_{\mathbf{V}}(R^{\mathbf{E}}_{\mu\nu}\, R^{\mathbf{E}\mu\nu})$ and then effect the replacement $\mathbf{g}_0 \to -\frac{1}{3}\,\mathbf{f}$.

Proof: Since modular adjustment concerns only the quark contribution, we need only consider the latter. Now (81) is rewritten, leaving $\mathbf{g}_0$ as it stands:

$$(81a)\quad i\,R^{\mathbf{E}}_{q\mu\nu} = \begin{pmatrix} (\mathbf{g}^0_{\mu\nu}-\mathbf{f}_{\mu\nu})\otimes\mathbb{1}_N & 0 & 0 \\ 0 & (\mathbf{g}^0_{\mu\nu}+\mathbf{f}_{\mu\nu})\otimes\mathbb{1}_N & 0 \\ 0 & 0 & (\mathbf{h}^{\cdot}{}_{\cdot\mu\nu}-\delta^{\cdot}_{\cdot}\mathbf{g}^0_{\mu\nu})\otimes\mathbb{1}_N \end{pmatrix} \otimes \mathbb{1}_3$$

($u_R$, $d_R$, $u_L\ d_L$; $u_R$, $d_R$, $u_L\ d_L$)

$$+\ \mathbf{g}^a_{\mu\nu}\, \mathbb{1}_4 \otimes \mathbb{1}_N \otimes \frac{\lambda_a}{2}\ .$$

The prescription (85) then yields:

$$(95)\ -tr_{\mathbf{V}_q}(R_{q\mu\nu}\, R^{q\mu\nu}) = 3N\,[(\mathbf{g}^0_{\mu\nu} - \mathbf{f}_{\mu\nu})(\mathbf{g}^{0\mu\nu} - \mathbf{f}^{\mu\nu}) + (\mathbf{g}^0_{\mu\nu} + \mathbf{f}_{\mu\nu})(\mathbf{g}^{0\mu\nu} + \mathbf{f}^{\mu\nu})$$
$$+ (\mathbf{h}^i{}_{k\mu\nu} - \delta^i{}_k\, \mathbf{g}^0_{\mu\nu})(\mathbf{h}^i{}_k{}^{\mu\nu} - \delta^i{}_k\, \mathbf{g}^{0\mu\nu})] + 2N\, \mathbf{g}^a_{\mu\nu}\, \mathbf{g}_a{}^{\mu\nu}$$

$$= 3N\,[(4\, \mathbf{g}^0_{\mu\nu}\, \mathbf{g}^{0\mu\nu} + 2\, \mathbf{f}_{\mu\nu}\, \mathbf{f}^{\mu\nu} + \mathbf{h}^i{}_{k\mu\nu}\, \mathbf{h}^i{}_k{}^{\mu\nu}] + 2\,N\, \mathbf{g}^a_{\mu\nu}\, \mathbf{g}_a{}^{\mu\nu}$$

$$= 3N\,[(4\, \mathbf{g}^0_{\mu\nu}\, \mathbf{g}^{0\mu\nu} + 2\, \mathbf{f}_{\mu\nu}\, \mathbf{f}^{\mu\nu} + \frac{1}{2}\mathbf{h}^s{}_{\mu\nu}\, \mathbf{h}_s{}^{\mu\nu}] + 2\,N\, \mathbf{g}^a_{\mu\nu}\, \mathbf{g}_a{}^{\mu\nu},$$

which the replacement $\mathbf{g}_0 \to -\frac{1}{3}\,\mathbf{f}$ changes into:

---

[9] up to the change $z^2 \to \mathrm{Tr}(\,|k_0^d|^4 + |k_0^u|^4 + \frac{1}{3}\,|k_0^e|^4\,)$ in [6] which we suggest.

(96) $3N\,[(\frac{4}{9} + 2)\, \mathbf{f}_{\mu\nu}\, \mathbf{f}^{\mu\nu} + \frac{1}{2} \mathbf{h}^{s}{}_{\mu\nu}\, \mathbf{h}_{s}{}^{\mu\nu}] + 2\, N\, \mathbf{g}^{a}{}_{\mu\nu}\, \mathbf{g}_{a}{}^{\mu\nu}$

$$= \frac{22}{3}\, N\, \mathbf{f}_{\mu\nu}\, \mathbf{f}^{\mu\nu} + \frac{3}{2}\, N\, \mathbf{h}^{s}{}_{\mu\nu}\, \mathbf{h}_{s}{}^{\mu\nu} + 2N\, \mathbf{g}^{a}{}_{\mu\nu}\, \mathbf{g}_{a}{}^{\mu\nu}.$$

**[14]** ***Remark***. The sum of the surface terms which we discarded in the expression of the action is the following:

(97) $$11\, \pi^2 \chi_4 + \frac{3}{8} \Delta \mathbf{s} + \frac{4}{3} A\, \Delta(|\Phi|^2),$$

where $\chi_4$ dv is the Euler form.

Proof: All the surface-terms come from $a_4(x, \mathbb{D}_A{}^2)$. We should first recover the $\chi_4$-term, which according to (71) was associated to the the square of the Weyl tensor in the expression $88\, \pi^2 \chi_4 - 18\, \mathbf{C}^2 = 8\,(11\, \pi^2 \chi_4 - \frac{9}{4}\, \mathbf{C}^2)$: whence the first term of (97), by comparison with (75).

On the other hand, the surface-terms which we discarded from $(4\pi)^2\, a_4(x, \mathbb{D}_A{}^2)$ inside the bracket $\{\}$ sum up to:

(98) $$\frac{1}{360} \mathrm{tr}_X\{12\, \mathbf{s}_{;}{}^{\alpha}{}_{\alpha}\, \mathbb{1} - 60\, E_{;}{}^{\alpha}{}_{\alpha}\}$$
$$= \frac{1}{360} \Delta \mathrm{tr}_X\{-12\, \mathbf{s}\, \mathbb{1} + 60\, (\frac{1}{4}\mathbf{s}\, \mathbb{1} - \frac{1}{2} c(R^{\mathbf{E}}) + i\, c^{\mu}\, [\mathbb{V}_{\mu}, \mathbb{\Phi}] + \mathbb{\Phi}^2)\}$$
$$= \frac{1}{360} \Delta \mathrm{tr}_X\{3\, \mathbf{s}\, \mathbb{1} + 60\, \mathbb{\Phi}^2\} = \frac{1}{360} \Delta\{3\, \mathbf{s}\, \mathrm{rank}\mathbf{V} + 60 \cdot 8\, A\, |\Phi|^2\}$$
$$= \frac{3}{8} \Delta \mathbf{s} + \frac{4}{3} A\, \Delta(|\Phi|^2),$$

where we used (32).

**[15]** ***Conclusion***. The axioms of noncommutative geometry produce the Euclidean bosonic Lagrangian from the fermionic Lagrangian as input. Apart from surface terms, this bosonic Lagrangian consists of the complete Einstein-Hilbert-Yang-Mills-Higgs Lagrangian of the standard model, a cosmological constant, a higher derivative gravity term of pure Weyl square form and the conformal curvature-Higgs coupling. After a proper normalization of the kinetic terms of the spin 2, spin 1 and spin 0 fields, all the parameters of the standard model can be read from equations (92, 93):

(99) $$m^2_{\mathrm{Planck}} = \frac{16\pi}{3K}\,(\,45 - \frac{2\,A^2}{B}\,)\,\Lambda^2\,,$$

(100) $$m^2_{\mathrm{Higgs}} = 4\,\Lambda^2\,,$$

(101) $$m_W^2 = \frac{A^2}{8B} \Lambda^2 ,$$

(102) $$g_2 = g_3 = \frac{1}{4} \sqrt{K} ,$$

(103) $$\sin^2\theta_w = \frac{3}{8} .$$

The cosmological constant is

(104) $$\frac{2}{K} ( 45 - \frac{2 A^2}{B} ) \Lambda^4$$

and the coefficient of Weyl term to the square is $- \frac{9}{4K}$ . The overall normalization constant is $K = 16 \pi^2$ as can be checked by considering the usual Dirac operator on the flat 4-torus with unit radii. There $a_0 = 4 (4\pi)^{-2}$ and all higher a's vanish. Therefore the Chamsedinne-Connes action is $2 \pi^2 \Lambda^4$ and coincides, for large $\Lambda$, with the number of eigenvalues smaller than $\Lambda$.

The conclusions (99-104) are evidently irrealistic. Chamsedinne and Connes argue that the renormalization group flow improves the situation. This line of reasoning meets the problem that fourth order gravity is renormalizable only in presence of a term in the curvature scalar squared [8] not found in $a_4$ (neither in $a_6$). Even ignoring gravity, $g_2 = \pi$ is a serious problem, because the hypothesis of the big desert yields $g_2 = 0.54 \pm 0.02$ at unification scale.

The conceptual beauty and unprecedented unification power of the Chamseddine-Connes action make us hope that these numerical problems will evaporate in the future.

## Index of notations

| | |
|---|---|
| $\mathcal{A} = \mathbb{A} \otimes \mathbf{A} = C^\infty(\mathbf{M}) \otimes (\mathbb{C} \oplus \mathbb{H} \oplus M_3(\mathbb{C}))$ | total algebra |
| $\mathbf{M}$ | space-time |
| $\mathbb{A} = C^\infty(\mathbf{M})$ | space-time algebra |
| $\gamma^\mu = \gamma^{\mu-1} = \gamma^{\mu *}$ | Dirac matrices |
| $\mathbb{S}_\mathbf{M}$ | spin bundle |
| $\mathbb{S}(\mathbf{M})$ | its module of smooth sections |
| $\mathbb{H}$ | algebra of quaternions |
| $\mathbf{A} = \mathbb{C} \oplus \mathbb{H} \oplus M_3(\mathbb{C})$ | internal algebra |
| $\mathbb{H} = L^2(\mathbb{S}_\mathbf{M}) \otimes H = \mathbb{H}_{part} \oplus \mathbb{H}_{antipart}$ | total Hilbert space |
| $\mathbb{E} = \mathbb{S}(\mathbf{M}) \otimes H = \mathbb{S}(\mathbf{M}) \otimes_\mathbb{A} \mathbf{E}$, $\mathbf{E} = \mathbb{A} \otimes_\mathbb{A} H$ | twisted bundle (dense subspace of $\mathbb{H}$) |
| $L^2(\mathbb{S}_\mathbf{M})$ | square integrable spinors on space-time |
| $H = \underline{H} \oplus \overline{\underline{H}}$ , dim $H = 90$ | internal Hilbert space of particles and antiparticles |
| $\underline{H} = \underline{H}_q \oplus \underline{H}_l$, dim $\underline{H}_q = 36$, dim $\underline{H}_l = 9$ | q for quarks and l for leptons |
| $\mathbb{D} = D \otimes id_H + \gamma^5 \otimes D$ | total Dirac operator, self adjoint |
| D | classical Dirac operator on $\mathbb{S}_\mathbf{M}$ |
| $D = \underline{D} \oplus \overline{\underline{D}}$ | internal Dirac operator of particles and antiparticles |
| $\underline{D} = \underline{D}_q \oplus \underline{D}_l$ | quarks and leptons |
| $\mathbb{D}_A = \mathbb{D} + A + \mathbb{J}A\mathbb{J}$ | covariant Dirac operator restricted to particles |
| A | gauge bosons and Higgs scalars represented on $H$ |
| $\nabla^\mathbf{M}$ | Levi-Civita connexion on $\mathbf{M}$ |
| $\mathbf{R}$ | its Riemann curvature tensor |
| $\mathbf{r}$ | Ricci tensor |
| $\mathbf{s}$ | scalar curvature |
| $\mathbf{C}$ | Weyl tensor |
| $\tilde{\nabla}$ | spin connexion |
| $\tilde{R}$ | curvature of $\tilde{\nabla}$ |
| $\nabla^\mathbf{E}$ | connexion on $\mathbf{E}$ parameterized by Hermitian (!) |
| $\mathbf{a}$ | U(1) vector potential, |
| $\mathbf{b}$ | SU(2) vector potential |
| $\mathbf{c}$ | SU(3) vector potential |
| $R^\mathbf{E}$ | curvature of $\nabla^\mathbf{E}$ parameterized by Hermitian (!) |
| $\mathbf{f}$ | U(1) curvature |
| $\mathbf{h}$ | SU(2) curvature |
| $\mathbf{g}$ | SU(3) curvature |
| $R'^\mathbf{E}$ | $R^\mathbf{E}$ transformed by unimodularity condition |
| $\mathbb{V} = \tilde{\nabla} \otimes id_\mathbf{E} + id_{\mathbb{S}(\mathbf{M})} \otimes \nabla^\mathbf{E}$ | connexion on $\mathbb{E}$ |
| $\mathbb{R} = \tilde{R} \otimes id_\mathbb{E} + id_{\mathbb{S}(\mathbf{M})} \otimes R^\mathbf{E}$ | its curvature |
| $\Phi_i = \bar{\Phi}^i$, i=1,2 | complex doublet of Higgs scalars |
| $\mathbb{\Phi}$ | Higgs scalars represented on $\underline{H}$ |

## Bibliography

[1] A. Connes, Non commutative geometry and reality, Journ. Math. Phys. 36, 6194 (1995)
[2] P.B. Gilkey, Invariance theory, the heat equation and the Atiyah-Singer index theorem, CRC Press (1995)
[3] D. Kastler, The Dirac operator and gravitation, Comm. Math. Phys. 166, 633 (1995), W. Kalau and M. Walze, Gravity, non-commutative geometry and the Wodczicki residue, Journ. Geom. Phys. 16, 327 (1995)
[4] A. Connes, Gravity coupled with matter and the foundation of non commutative geometry, hep-th/9603053
[5] A. Connes, Brisure de symmétrie spontanée et géométrie du point de vue spectral, Exposé Bourbaki, June 1996
[6] A. Chamseddine and A. Connes, The spectral action principle, hep-th/9606001, A universal action formula, hep-th /9606056
[7] B. Iochum, D. Kastler and T. Schücker, Spectral model and fuzzy mass relations, Contemporary Math. to appear (1995), A detailed account of Alain Connes' version of the standard model in non-commutative differential geometry.V. Spectral model and fuzzy mass relations, (1996) preprint CPT.
[8] K.S. Stelle, Renormalization of high-derivative quantum gravity, Phys. Review 16, 953 (1977)